\pdfoutput=1
 
\documentclass[useAMS,usenatbib]{mn2e}
\usepackage{graphicx}
\usepackage{amssymb,amsmath}
\usepackage{threeparttable}

\title[Deep LFs \& CMRs ]{Deep Luminosity Functions and Colour-Magnitude Relations for 
Cluster Galaxies at $0.2 < z < 0.6 $}

\author[R. De Propris, S. Phillipps \& M. Bremer]  {
R. De Propris$^1$,  S. Phillipps$^2$ and M. Bremer$^2$\\
   $^1$ European Southern Observatory, 3107 Alonso de Cordova, Santiago, Chile\\
   $^2$H. H. Wills Physics Laboratory, University of Bristol, Tyndall Avenue, Bristol, BS8 1TL, United Kingdom
   }

\begin{document}

\date{}

\pagerange{\pageref{firstpage}--\pageref{lastpage}} \pubyear{2012}

\maketitle

\label{firstpage}

\begin{abstract}

We  derive deep $I$ band luminosity functions and colour-magnitude diagrams from  HST imaging for eleven
$0.2<z<0.6$ clusters observed at various stages of merging, and a comparison sample of five more relaxed
clusters at similar redshifts. The characteristic magnitude $M^*$ evolves passively out to $z=0.6$, while the
faint end slope of the luminosity function is $\alpha \sim -1$ at all redshifts. Cluster galaxies must have 
been completely assembled down to $M_I \sim -18$ out to $z=0.6$.  We observe tight colour-magnitude
relations over a luminosity range of up to 8 magnitudes, consistent with the passive evolution of ancient 
stellar populations. This is found in all clusters, irrespective of their dynamical status (involved in a collision
or not, or even within subclusters for the same object) and suggests that environment does not have a 
strong influence on galaxy properties. A red sequence luminosity function can be followed to the limits of  
our photometry: we see no evidence of a weakening of the red sequence to $z=0.6$. The blue galaxy fraction
rises with redshift, especially  at  fainter absolute magnitudes.  We observe bright blue galaxies in clusters at
$z > 0.4$ that are not encountered locally. Surface brightness selection  effects preferentially influence the
detectability of faint red galaxies, accounting for claims of evolution at the faint end.

\end{abstract}


\section{Introduction}
Galaxy populations in clusters may be regarded as a volume-limited sample of objects, observed 
at the same cosmic epoch and lying within similar peaks in the dark matter distribution at each 
lookback time. Cluster members have a high surface density on the sky and can therefore be 
identified (in a statistical sense) from the surrounding field, allowing us to study the properties 
of galaxies (even at high redshift, with {\it bona fide} clusters now known at $z \sim 2$; 
\citealt{Zeimann2012}) without observationally expensive spectroscopic campaigns. The evolution
of cluster galaxies provides a benchmark to test theories of galaxy formation and especially the
relative influence of initial conditions versus environmental effects. Mechanisms such as ram stripping 
by the hot X-ray gas, multiple interactions between galaxies (harassment) and tides induced by the 
cluster potential may all conspire to alter the evolution of galaxies in clusters. Nevertheless, these 
objects still provide useful clues to the history of galaxy assembly and the formation of their stellar 
populations. We may envisage  that  through studies of clusters of different masses, and lower density 
regions within individual objects (e.g., outskirts, subclusters) we will be able to relate cluster galaxy
evolution to the  more general case of field galaxies, while a comparison between the behaviour in the field 
and clusters may yield experimental tests of how galaxy evolution depends on their surroundings.

A series of papers have examined the luminosity function of galaxies in clusters to trace the
history of mass assembly and these have generally agreed that most massive cluster galaxies 
have formed rapidly at high redshift (e.g., \citealt{DePropris1999, Andreon2006a, DePropris2007,
Muzzin2008,Mancone2010} and references therein). Local and high redshift cluster galaxies also
exhibit a well-defined `red sequence' of early-type galaxies. The red sequence is observed even in 
the most distant clusters yet studied in detail (e.g., \citealt{Papovich2010}) and is believed to be
driven by a mass-metallicity relation, with the small intrinsic scatter representing a small spread in
ages. Together with the tight Fundamental Plane relations, even at high redshifts \citep{Holden2010}, 
this is consistent with the pure passive evolution of the stellar populations formed at high redshift in
short star formation episodes. 

While massive galaxies may have formed early, it is likely that low luminosity (dwarf) galaxies have 
undergone a more extended formation history, as in the `downsizing' model of \cite{Cowie1996}. For
example, Local Group dwarfs have complex stellar populations, with multiple star formation and 
enrichment episodes (e.g., \citealt{Weisz2011}). In particular, star-forming dwarf irregulars may be
efficiently quenched to produce quiescent dwarf spheroidals, with this process moving progressively
to higher masses with increasing redshift (\citealt{Cowie1996, PerezGonzalez2008}). In agreement with this 
picture, the luminosity function of red sequence galaxies appears to weaken at the faint end as a function of 
redshift in the COSMOS and Extended Groth Strip fields (\citealt{Bell2004, Faber2007} {\it et seq.}). 
A similar decrease in the fraction of red sequence galaxies may be taking place in clusters as well
(\citealt{DeLucia2007} and subsequent studies, but see \citealt{Andreon2008} for an opposing view). 
However, \cite{Mancone2012} determine the faint end slope $\alpha$ of the luminosity function in 
the rest-frame $H$ band for seven $z \sim 1.5$ clusters and find that it is consistent with the local value 
of $\alpha \sim -1$, which implies an early formation history for these low luminosity objects as well.

In order to explore complex themes, such as these, it is often useful to examine possible extreme cases. 
In particular, one of the proposed mechanisms for the quenching of star formation in cluster galaxies is
ram pressure stripping as galaxies move through the cluster gas (e.g., Quilis, Moore \& Bower 2000). The
most extreme such cases are collisions between clusters, as for instance in the `Bullet' cluster (1E0657-558;
\citealt{Tucker1998,Markevitch2002}), where X-ray observations show a bullet-like cloud which has passed
through the main cluster at a velocity estimated at $\sim 4500$ km s$^{-1}$ around 150 Myr ago. Combinations
of optical, X-ray and gravitational weak lensing maps demonstrate that the shocked, colliding gas has been
swept out of the clusters and is now situated between the outward travelling galaxies and dark mass concentrations
(see e.g., \citealt{Markevitch2004, Clowe2006}). This is equivalent to a ram stripping wind at least two orders of
magnitude greater than has been experienced by any galaxy in its motion through its own cluster X-ray gas. In
these environments, the effects of ram stripping should trump any other influence on the evolution of cluster members
and can be studied in relative isolation.

In the present paper, we therefore consider the  effect that a major, supersonic, collision between clusters has 
had on the evolution of the galaxies they contain. Besides being an extreme dynamical environment, there may 
in this case also be the opportunity to look at the individual influences of the gas, galaxies and dark matter. 
This provides a useful counterpart to the similar study of more `normal' clusters that we have initially carried 
out in \cite{Harsono2009} and a few comparison objects that we present here as part of a broader analysis.

The following sections describe, in this order, the dataset and its analysis, the  luminosity functions for 
the whole population (split by  subcluster where possible), the colour-magnitude relations and from these 
the luminosity distributions for galaxies in the red sequence and blue cloud. We finally discuss our 
findings in context and  present avenues for future work. We assume the latest cosmological parameters from 
the WMAP 9 year dataset presented in \cite{Hinshaw2012}. Extinction values for our fields are derived from the 
latest reanalysis of COBE/DIRBE data by \cite{Schlafly2011}.

\section{Dataset}

The clusters selected for this study consist of a sample of 11 `collisional' clusters, identified
on the basis of deep X-ray observations, optical data and analysis of the weak and strong
lensing. In addition, our sample includes a few `normal' clusters at similar redshifts, which are 
part of a broader study we are carrying out. These are discussed separately, although the analysis 
we describe below also applies to these latter objects.

\begin{figure*}
\includegraphics[height=8in]{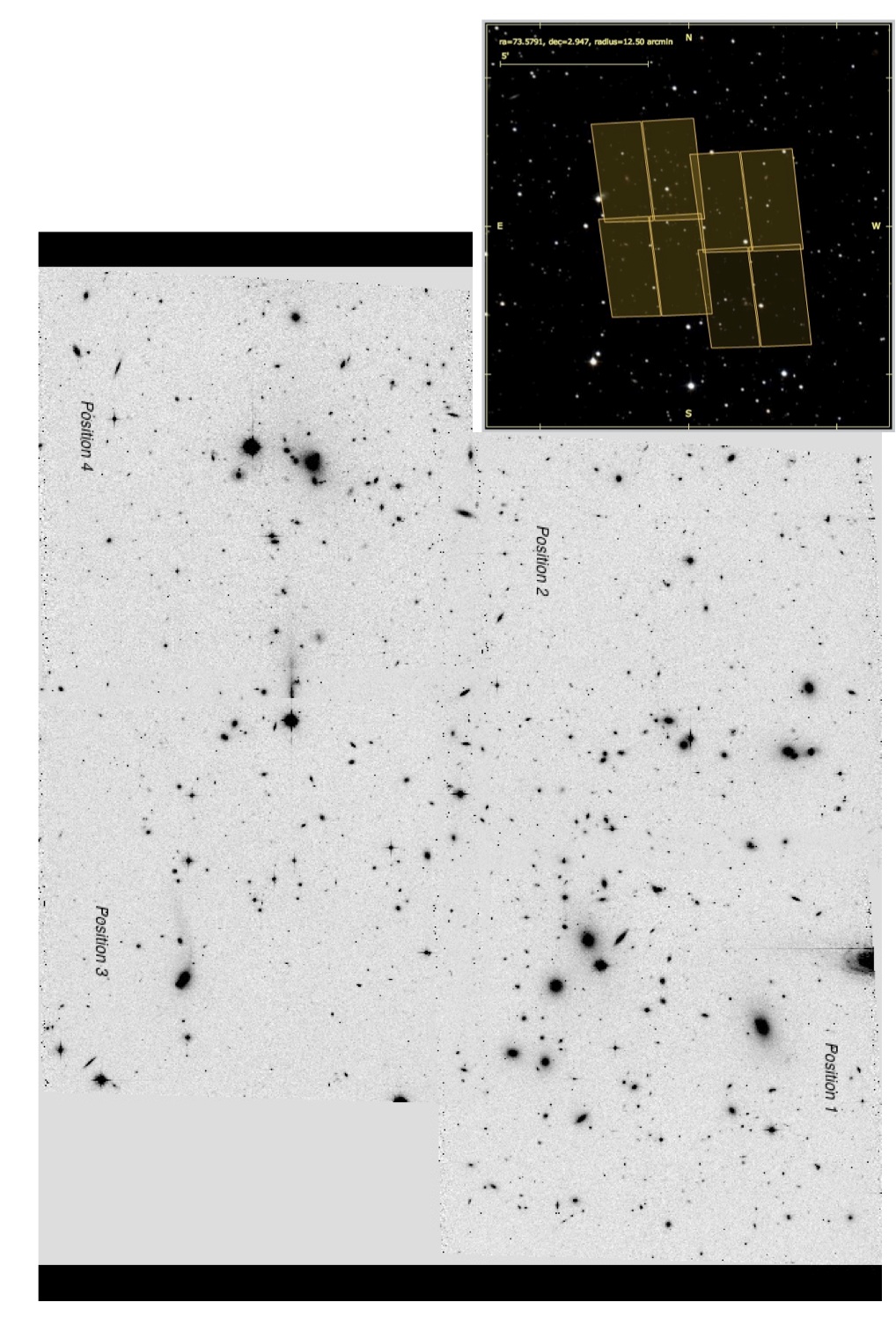}
\caption{Mosaicked $I$ band image from HST showing the available data for Abell 520. In this
figure, North is up and East is to the left.  The positions referred to in Table~\ref{clus} are
marked in the figure. The inset shows the positions of the HST images on the sky, with the
ACS fields of view in semi-opaque yellow. The background image is from the Palomar Observatory
Sky Survey.}
\label{a520foot}
\end{figure*}

We have retrieved $V$ (F606W) and $I$ (F814W) images of our selected objects from the Hubble 
Legacy Archive (HLA). Table 1 presents  a summary of the images, exposure times, program 
IDs and other necessary information. In some cases we have photometry for other bands and this will 
be considered in future papers and in other contexts. All data were retrieved as fully calibrated and  
drizzled files from the HLA server and/or the Barbara Mikulski Archive for the Space Telescope (MAST). 
Grayscale images of these objects, from the Palomar Observatory Sky Survey (POSS), with the positions 
of ACS fields superposed, can be found in the Appendix section. Here we show the one for Abell 520 
in Figure~\ref{a520foot} by way of example.

Our sample includes the following clusters:

\begin{itemize}

\item Abell 520  is  arguably the most primitive object and consists of three separate groups accreting 
along filaments to form a massive cluster \citep{Girardi2008}. The X-ray plasma appears to be 
separate from the galaxies, which instead coincide with the dark matter distribution \citep{Clowe2012}. 
The data here consist of four ACS fields in $V$ and $I$ (see Table~\ref{clus} for a summary) in a
$2 \times 2$ mosaic with a slight overlap. Figure~\ref{a520foot} shows the ACS footprints over a 
Palomar Observatory Sky Survey plate. We refer to these as Positions 1-4 in a 'Z' pattern, with  
position 1 at the top right. In order to provide an idea of the quality of the data, we also show an
HST mosaic for this object in Figure~\ref{a520foot}. Similar figures (showing the HST footprints, but
not the HST images, which can be easily retrieved from the archives) are presented for all other clusters 
we consider in this paper in the Appendix.

\item Abell 1758 contains two subclusters, each of which seems to be undergoing a separate 
merger. The Northern cluster (studied here) has a double peaked X-ray structure, of which one 
component coincides with the galaxy distribution \citep{Ragozzine2012}. Two ACS fields have
been observed, one on each peak in the Northern cluster. Position 1 in Table~\ref{clus} is to
the South-East and Position 2 to the North West.

\item Abell 2163 also has complex dynamics and exceptional X-ray properties (high temperature and 
luminosity) and is believed to be a multiple merger observed 1 Gyr after the main crossover 
\citep{Bourdin2011, Soucail2012}. Two ACS pointings are available in this field. Position 1 
in Table~\ref{clus} is to the North East and position 2 to the South West.

\item The Bullet cluster has been discussed in detail above. The available HST pointings image both 
the main cluster to the East and the bow shock region corresponding to a lower mass subcluster to 
the West that has crossed the more massive object about 0.2 Gyr ago.

\item Abell 2744 appears to have a very complex structure, with `dark', `ghost', `bullet' and `stripped' 
substructures \citep{Merten2011}, and may be a very active cluster merger. \cite{Owers2011} have used radial 
velocity information to trace the paths of three merger components (two major and one minor) in this 
cluster. We use a single HST image in this cluster, covering the main structures.

\item MACS0553.4-3342 lies at $z=0.407$, is very luminous in the X-rays and seems to consist of 
a linear collision between nearly equal mass subclusters \citep{Ebeling2010}. 

\item MACS J0358.8-2955 is a massive cluster merger at  $z=0.434$ where Chandra images show a 
strong separation between dark matter and the X-ray  gas with a linear post-collision geometry. It appears
 to be the results of a complex merger of at least three subclusters \citep{Hsu2012}.  
 
\item MACS1226.8+2153 is a rare triple cluster merger at  $z=0.43$ lying in a deep node of filaments and 
dark matter structures, showing several large arcs \citep{Ebeling2010}. The three regions are denominated
in Table~\ref{clus} as the Central, North Eastern and Southern components and these can also be
identified from the associated POSS images with the ACS coverage in the Appendix.
 
\item DLSCL J0916.2+2951 is a  dissociative cluster merger at $z=0.53$ observed 
about 0.7 Gyr after first pass and may represent a more evolved version of the Bullet Cluster  
\citep{Dawson2012}. The two components (see Appendix for the HST coverage) are referred
here as the Southern and Western one.

\item MACS0717+3745 at $z=0.55$ is a very massive object, within a large filament, 
and consisting of an active triple merger with very complex dynamics \citep{Ma2009,Limousin2012}. 

\item CL 0025-1222 consists of two merging subclusters of nearly equal mass at   $z=0.58$, with the  dark 
matter distribution coinciding with the galaxies and clearly separated from the X-ray emitting gas 
\citep{Bradac2008}.

\end{itemize}

\begin{table*}
 \centering
 \begin{center}
 \begin{minipage}{0.75\textwidth}
  \caption{`Bullet-like' clusters studied in this paper}
  \begin{threeparttable}
  \begin{tabular}{lcccccc}
  \hline \hline
   Cluster & $z$ & Passband & Exposure &  Area  & Proposal & PI \\
               &        &                &         [s]            &       [arcmin$^2$]\\
 \hline
A520\tnote{a}  & 0.199 & $V$ & 2332 &  45.19 & 12253 & Clowe \\
                              &           &  $I$ & 4570 &            &              &   \\
A2163\tnote{b} & 0.203 & $V$ & 2340 &  23.33         &  12253 & Clowe \\
                                &           & $I$ & 4596  &                   &                &             \\
A1758\tnote{b} & 0.279 & $V$ & 2544 & 12.42           &  12253  & Clowe \\
                               &           & $I$ & 2500 & 12.40           &                &            \\
Bullet\tnote{b} & 0.296 & $V$ & 2336 & 18.06  & 10200 & Jones \\
                      &           & $I$ & 4004 & 11.33\tnote{c}  & 10200 & Jones \\
                      &          &  $I$ & 4480 & 11.33\tnote{d} & 11491\tnote{e} & Kneib    \\
A2744               & 0.308 &  $V$ & 5356 & 12.05\tnote{f}     &   11689 & Dupke    \\
                           &            &  $I$ &             &       &                &               \\ 
MACS0553.4-3342 & $0.41$ & $V$ & 2092 & 11.70\tnote{g} & 12362 & Ebeling \\
                                    &               & $I$  & 4572 & 11.70 &              &                \\
MACS1226.8+2153 C\tnote{h} & $0.43$ & $V$ & 1200 & 11.88  & 12166 & Ebeling \\
                                                   &                & $I$                    & 1440 & 11.87  &              & \\
 MACS1226.8+2153 NE       &                 &$V$ & 2040 &   8.35 & 12368 & Morris \\
                                                   &                 &$I$                    & 2040 &   8.35  &              & \\
 MACS1226.8+2153 S          &                 & $V$ & 2040 & 12.05  & 12368 & Morris \\
                                                   &                & $I$                     & 2040 &  12.05 &               & \\                                                                
 MACS0358.8-2955 & $0.43$ & $V$ & 2120 & 11.70 & 12313 & Ebeling \\
                                     &              &    $I$ & 4620 & 11.70 & & \\
 J0916+2951 South & $0.53$ & $V$ & 2520 & 12.41 & 12377 & Dawson \\
                                     &               & $I$ & 4947 & 11.88 & & \\
 J0916+2951 West   & $0.53$ & $V$ & 2520 & 11.88 & 12377 & Dawson \\
                                     &               & $I$ & 4947 & 11.88 & & \\
 J0717+3745             & $0.55$ & $V$ &          &            &                  &        \\
CL0025-1222 & $0.58$ & $V_{F555W}$ & 4140 & 11.71 & 10703 & Ebeling \\
                          &               & $I$                     & 4200 & 11.71 &             &                 \\
CL0025-1222\tnote{i} & & $V_{F555W}$ & 4470 & 11.71 & 9722 & Ebeling \\
                                         &  & $I$                    & 4560 & 11.71 &           &                \\
 \hline

\end{tabular}
     \begin{tablenotes}
       \item[a] $2 \times 2$ mosaic covering Chandra image
       \item[b] 2 separate pointings on each concentration
       \item[c] Eastern Subcluster
       \item[d] Western Subcluster
        \item[e] Total $I$ coverage is 19.34 arcmin$^2$; combined $V$ + $I$ area is 16.28 arcmin$^2$
       \item[f] Combined $V$ and $I$ area is 10.21 arcmin$^2$
       \item[g] Combined $V$ and $I$ area is 9.13 arcmin$^2$
       \item[h] $V$ and$I$ overlap is 8.75 arcmin$^2$
       \item[i] Outer field
     \end{tablenotes}
  \end{threeparttable}
\end{minipage}
\end{center}
\label{clus}
\end{table*}

\section{Data Analysis: The Bullet Cluster}

\begin{figure}
\includegraphics[width=0.48\textwidth]{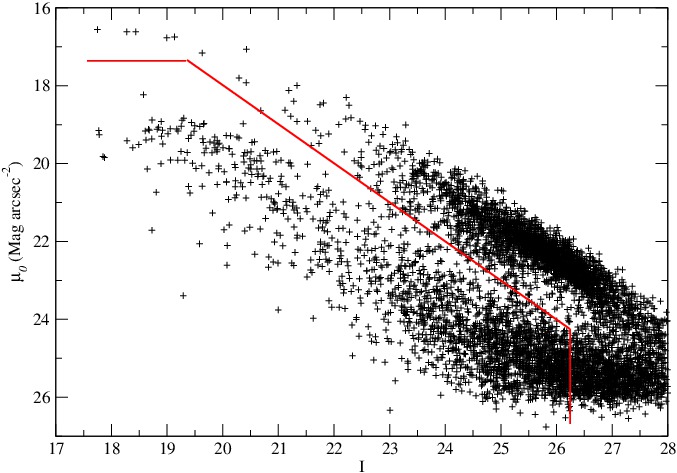}
\caption{Central surface brightness (in a small aperture, equivalent to the detection aperture
used by Sextractor) vs. total magnitude for all objects in the field of one of our targets. Stars
and other compact objects are easily separated as they follow a linear sequence. The thick
red line shows the surface brightness adopted to discriminate galaxies from stars as a function of
magnitude.}
\label{sgsep}
\end{figure}

\begin{figure}
\includegraphics[width=0.48\textwidth]{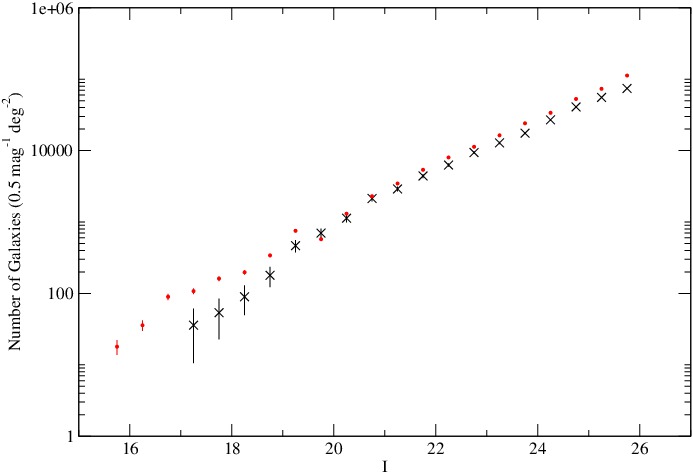}
\caption{$I$ band number counts for two areas in the CANDELS and GOODS surveys that we
have re-analysed so far. The black crosses refer to 106 sq. arcmin of the Extended Groth
Strip, while the red dots are for the 201 sq. arcmin of the CANDELS and GOODS images 
that we have studied. Error bars are Poissonian and do not include (here) clustering errors.}
\label{icts}
\end{figure}

\begin{figure}
\vspace{0.52cm}
\includegraphics[width=0.48\textwidth]{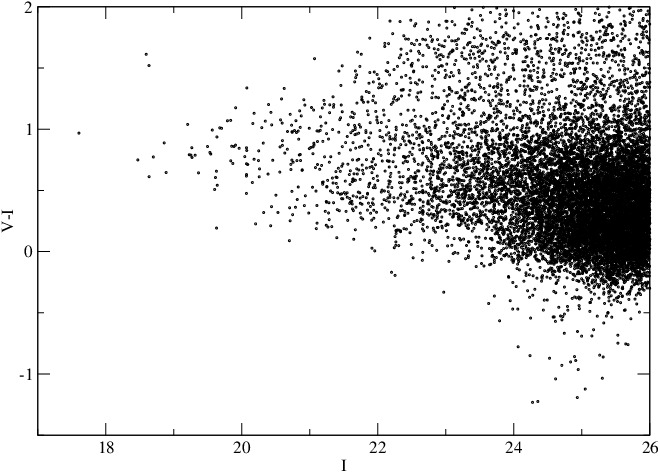}
\caption{$V-I$  vs $I$ colours for galaxies in one of the CANDELS/GOODS survey regions. The total 
$V$+$I$ field that we have analysed so far is 274 sq. arcmin.}
\label{icmd}
\end{figure}

For each cluster we ran Sextractor \citep{Bertin1996} with parameters as defined in
\cite{Harsono2009} to yield accurate and complete detections with a minimum of 
contamination from cosmetic features and especially arclets which are present in 
abundance in the cluster fields. We obtained both a `total' magnitude and an aperture
value, set to match the metric aperture of $\sim 5$ kpc (diameter) for Coma galaxies 
presented in \cite{Eisenhardt2007}. All magnitudes are in the AB system, using the latest 
zeropoints as calculated on the HST web site. The images were then visually inspected to 
remove spurious objects, arcs, bleeding trails from bright stars, detections on CCD edges, 
satellite streaks and other contaminants. Here we present our analysis of the Bullet cluster
as an example of the procedures we carried out on all objects in our sample.

We used  a comparison between the aperture magnitude (used for detection) and total magnitude 
to carry our star-galaxy separation, as shown in Figure~\ref{sgsep} for an example object. As we
can see in this figure, we are also complete, in surface brightness, to about $I=25$ where we see
a sharp cut in the central surface brightness distribution at $\sim \mu_I=25.5$ mag arcsec$^{-2}$.
Based on stellar counts and simulations, the actual image completeness is 100\% at $I=27$ and the
$5\sigma$ detection threshold is close to $I=28$, but galaxies are detected on the basis of their
central surface brightness, which is much lower than for a star. Below the $\mu_I=25.5$ mag arcsec$^{-2}$
limit galaxies do exit, but are not detected and therefore not measured; correcting for this incompleteness
requires knowledge of the surface brightness distribution of galaxies, which is poorly known (see discussion of
MS1358+62 below). We adopt $I=26$ as our photometric limit, where we are still highly complete and we
can adequately separate stars and galaxies on their basis of their central concentrations. However, it is
clear that at $I > 25$ the galaxy counts are incomplete because of surface brightness selection effects
(not the image detection limits), although we apply a similar surface brightness cut to the background
fields  (see below) we use, so that we are not favouring or disfavouring cluster members.

\begin{figure*}
\begin{tabular}{cc}
\includegraphics[width=0.5\textwidth]{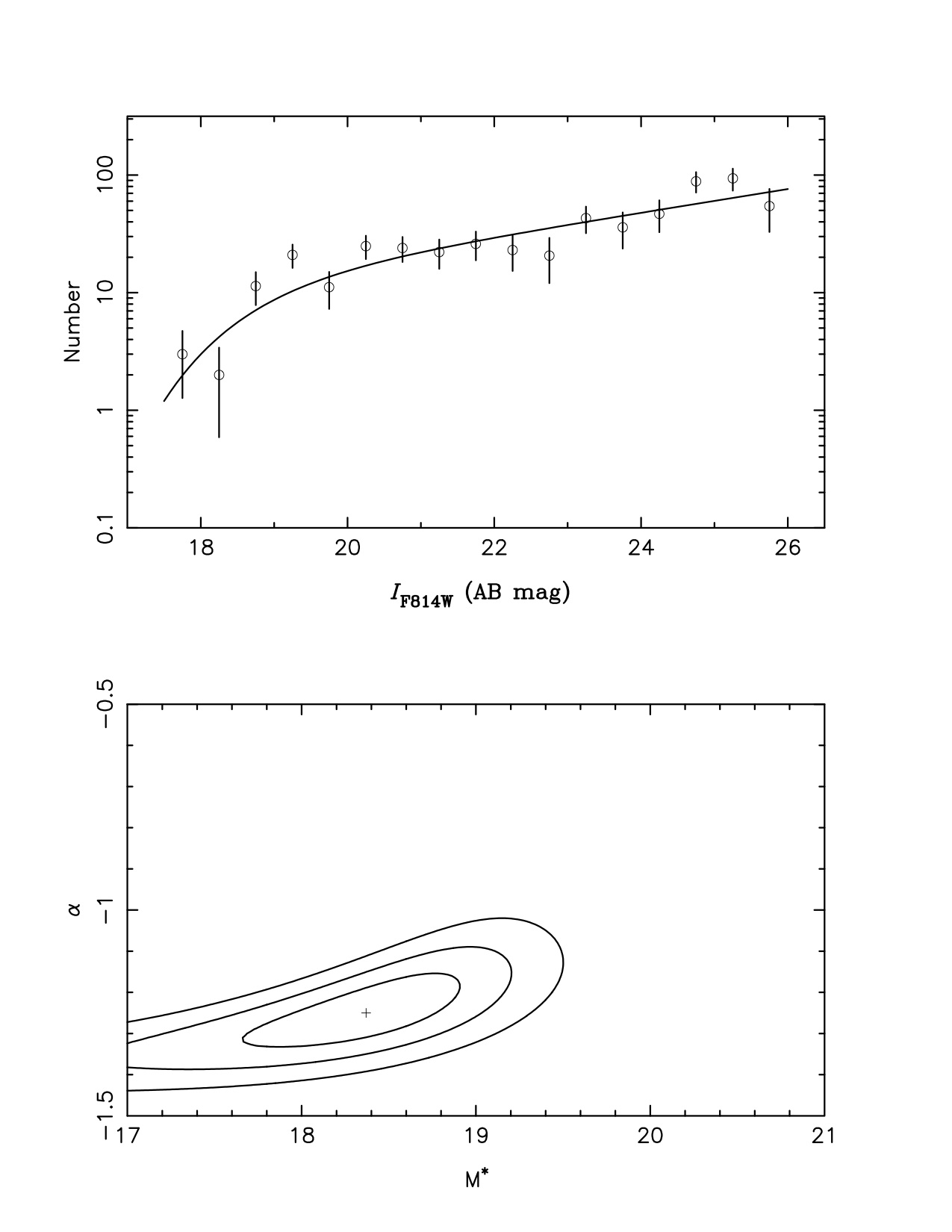} & \includegraphics[width=0.5\textwidth]{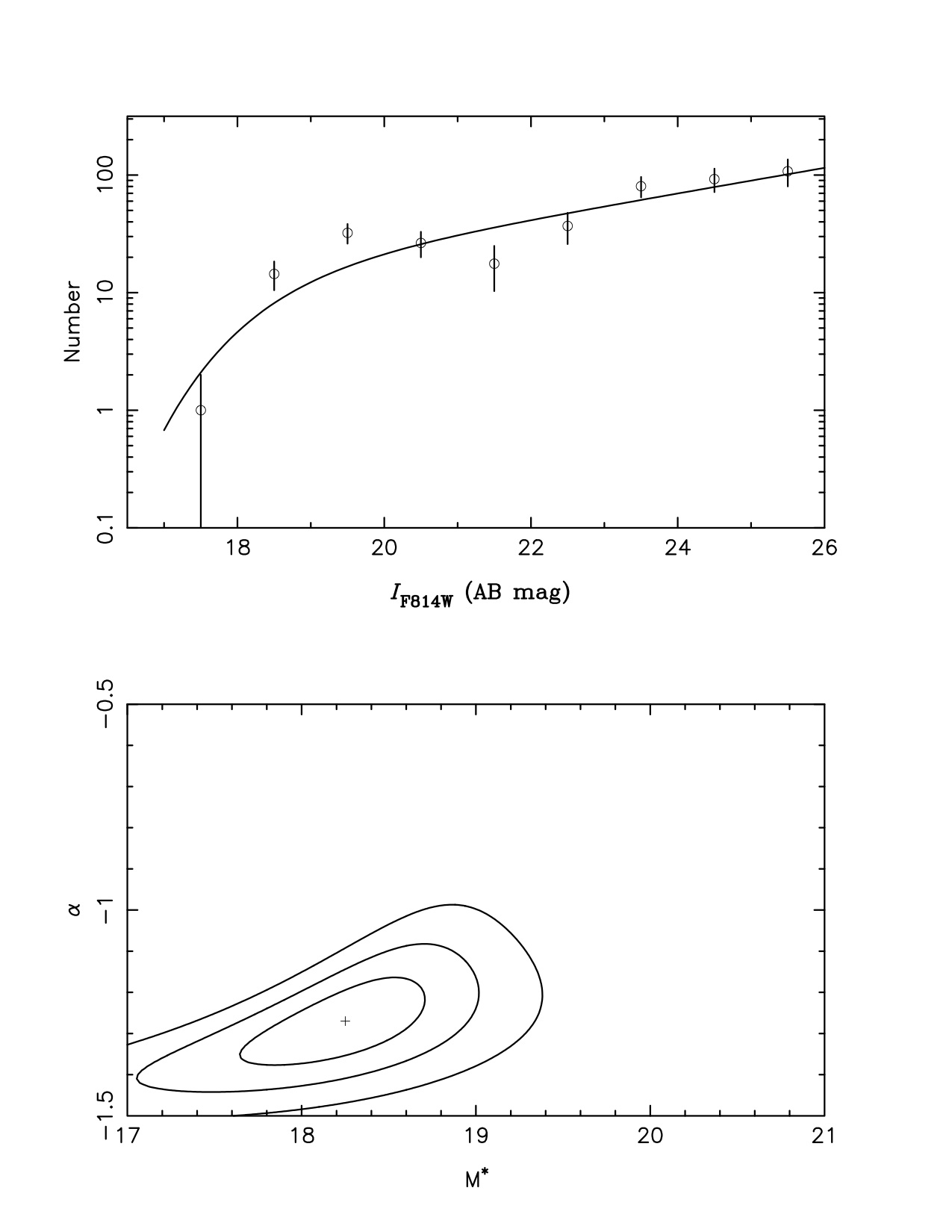} \\
\end{tabular}
\caption{Luminosity functions and best fits, with associated error ellipses, for galaxies in the Eastern
(left) and Western (right) subclusters of the Bullet Cluster.}
\label{bulletlfs}
\end{figure*}

Galaxies in the cluster fields consist of cluster members plus foreground and background objects
along the line of sight from Earth orbit to the cluster. We remove intervening galaxies statistically, 
by using counts in  publicly available fields imaged with HST, chiefly the GOODS \citep{Giavalisco2004} 
and CANDELS \citep{Grogin2011, Koekemoer2011} surveys, that have the required depth and
are wide enough to allow us to minimise the effects of clustering variations from intervening
large scale structures. These are analysed in the same fashion as the cluster fields (photometry 
with the same parameters followed by similar visual inspection and selection of targets). Number 
counts in $I$ for the  whole fields  analysed so far as well as the $I$ selected $V-I$ colour-magnitude 
distribution are shown in  Figure~\ref{icts} and Figure~\ref{icmd} (where we show one subfield only, for 
purposes of illustration).

For all clusters we scaled the galaxy number counts in the field to the areas covered and
subtracted the non-cluster contribution statistically, including contributions to the errors due
to clustering variations \citep{Peebles1975}, following the method of \cite{Huang1997} as
applied by \cite{Driver2003} in Abell 863 and \cite{Pracy2004} in Abell 2218. We should note 
here that the redshifts of our clusters are such that their distance is much greater than the largest 
structures observed in the 2dF and SDSS surveys and well beyond the maximum scale expected 
in CDM cosmology, and therefore that galaxy counts in their direction should approximate 
homogeneity. 

The subtracted counts yield a luminosity function for cluster galaxies (in a statistical sense), which we
fit with a standard Schechter (1976) function, using an {\tt amoeba}-like $\chi^2$ fitting package, that
also gives us error ellipses at the $1,2,3\ \sigma$ level. As an example of our approach we show here the
luminosity function in the $I$ band for the Bullet Cluster (both subclusters separately) and its best fit (with
the associated error ellipse) in Figure~\ref{bulletlfs}. Other LFs for individual clusters are shown in the 
Appendix (on-line only). The luminosity function of galaxies in the Bullet subclusters appears to be a 
reasonable fit to a single  Schechter function with parameters as given in Table 2. 

\begin{figure*}
\begin{tabular}{cc}
\includegraphics[width=0.5\textwidth]{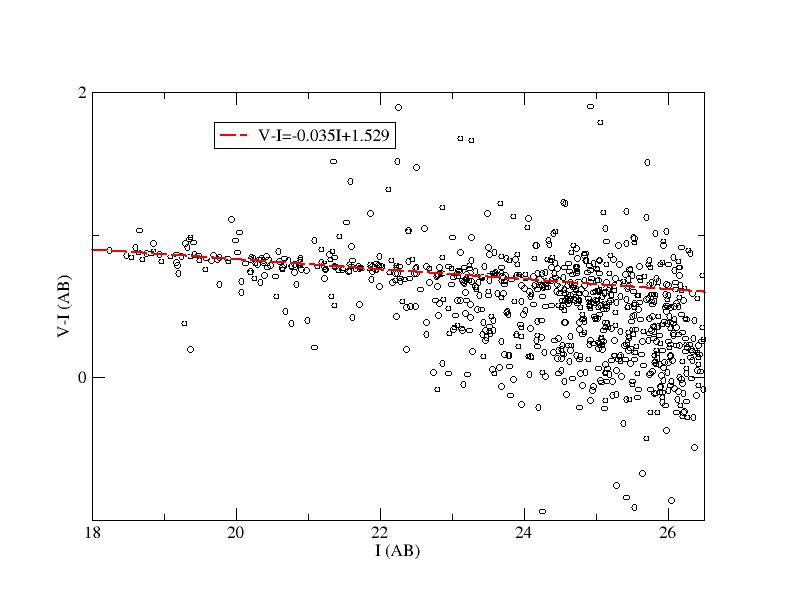} & \includegraphics[width=0.5\textwidth]{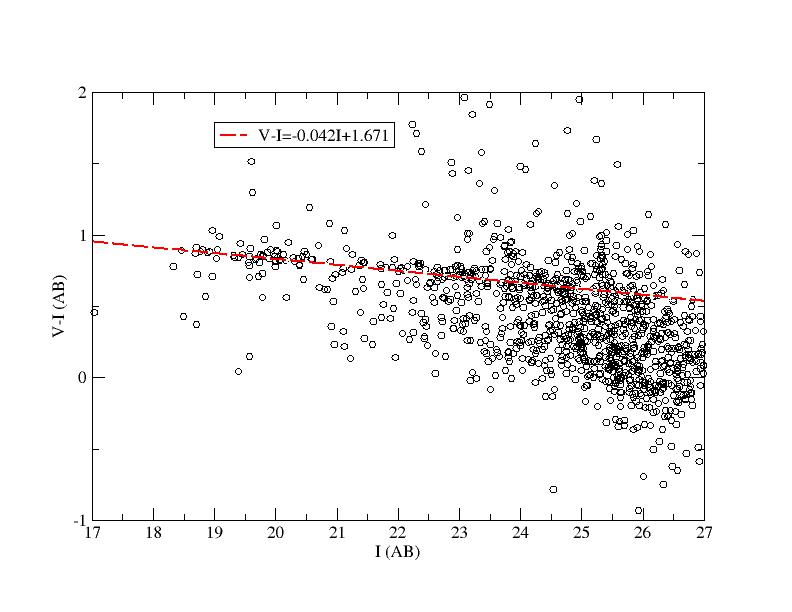} \\
\end{tabular}
\caption{Colour-magnitude relations and best fitting robust straight line, for galaxies in the Eastern
(left) and Western (right) subclusters of the Bullet Cluster.}
\label{bulletcmr}
\end{figure*}

We also derive a colour magnitude relation, by plotting $V-I$ colours (in fixed apertures) vs. total magnitude
and using a 'robust' routine to derive the best fitting straight line to the red sequence \citep{Armstrong1978}.
These are plotted, for the Bullet the colour-magnitude relations and best fits are plotted in Fig~\ref{bulletcmr},
while the slope and intercept of the relations are also shown in Table 2. 

From this we can assign galaxies to the red sequence and blue cloud, and derive independent luminosity
distributions for galaxies in these colour ranges. We plot the colour distribution relative to the red sequence
(where the colour of the sequence  itself is set to 0 at all $I$) for galaxies in the Bullet cluster in 
Figure~\ref{coldist}.  This shows a characteristic shape with a peak corresponding to the red sequence and 
a tail to  redder colours (mostly background galaxies but possibly including dusty cluster members) and 
a second peak for the blue cloud (which includes background galaxies). The two peaks can be separated 
roughly at a colour difference of $-0.24$ which we adopt here as a first order discriminant between 
red sequence and blue cloud objects. For example, \cite{DeLucia2007} use a colour range of $\pm 0.3$ mag.
around the red sequence as their definition. The blue objects here include a wider range of galaxies 
than those usually considered within the classical Butcher-Oemler effect \citep{Butcher1978,Butcher1984}, 
which are usually bluer than the red sequence by about 0.4 mag. (in $V-I$ colour at these redshifts), and 
therefore represent a more complete sampling of the population of recently star-forming galaxies in 
each cluster.

For each cluster, we carry out an equivalent colour-cut on the field galaxy distributions (these of course differ
from cluster to cluster) and then subtract the expected number counts for field galaxies within the colour 
regions corresponding to red and blue galaxies in each case. We show the colour distribution of the reference 
fields for the case of the Bullet cluster in Figure~\ref{coldist}, where we can graphically see the impact of
contamination on cluster membership (in a statistical sense) as a function of colour. We carry out a similar 
analysis for all other clusters in the sample. We then derive luminosity distributions for galaxies in the red 
sequence and blue cloud.

\begin{figure}
\includegraphics[width=0.5\textwidth]{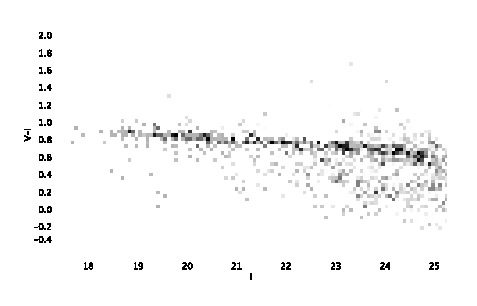} \\
\caption{Background subtracted colour-magnitude diagram for the Bullet Cluster, created 
as described in the text. The red sequence is prominent and thin to $I=25$ while blue cloud
galaxies become significant at $I > 22$, suggesting that most star formation takes place
among low-mass objects.}
\label{colprob}
\end{figure}

\begin{figure}
\includegraphics[width=0.5\textwidth]{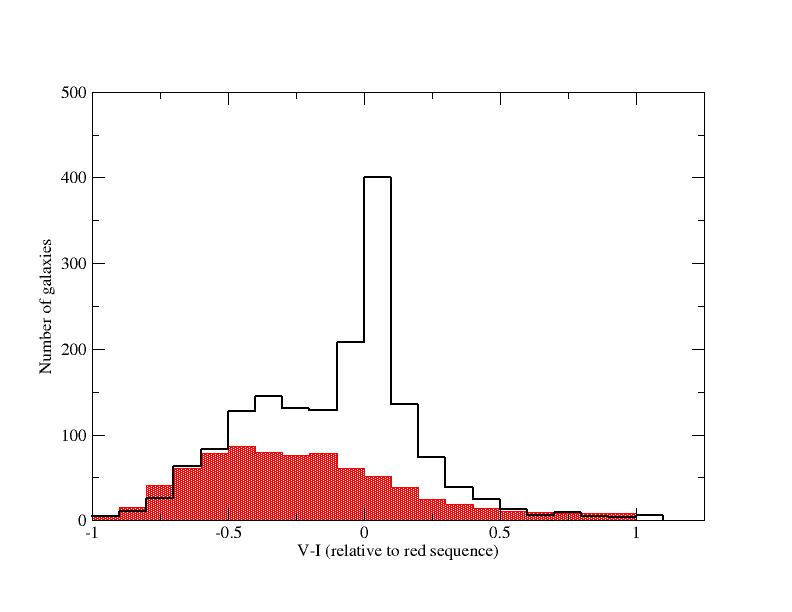} \\
\caption{Distribution of galaxies in colour in the Bullet cluster, for objects with $I < 26$, where the red sequence 
is set to have 0 colour (at each $I$ magnitude). The black histogram shows all galaxies in the cluster field of view, 
while the red histogram shows the normalised distribution for field galaxies.}
\label{coldist}
\end{figure}

The colour split is illustrated in a different way in Figure~\ref{colprob}. Each galaxy in the Bullet cluster field
has been weighted by its cluster membership `likelihood' and the density of weighted points in the 
colour-magnitude is shown. To obtain the likelihood (of being a cluster member) for any particular galaxy, we
determine the number of 'background field' (GOODS/CANDELS) galaxies ($B$) which have that galaxy as their 
nearest match in C-M space. If the background field has an area $A$ times the area of the cluster field then 
the likelihood that the chosen galaxy is a cluster member can be represented by $1 - B/A$. This can 
be negative, due to random fluctuations (though the sum of the weights gives the correct total number of 
cluster galaxies), so we smoooth slightly over the C-M distribution. Almost all galaxies redder than the red 
sequence are removed but the cluster's blue cloud is preserved.

\begin{figure*}
\begin{tabular}{cc}
\includegraphics[width=0.5\textwidth]{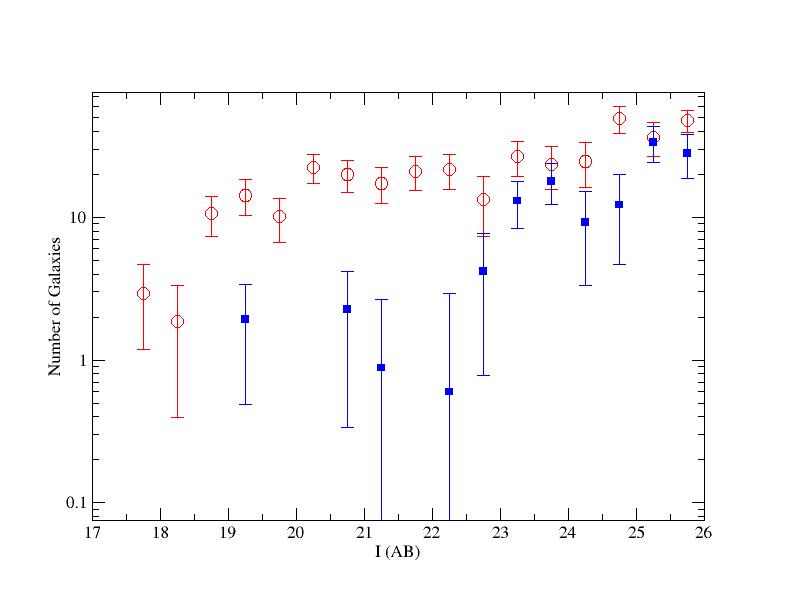} & \includegraphics[width=0.5\textwidth]{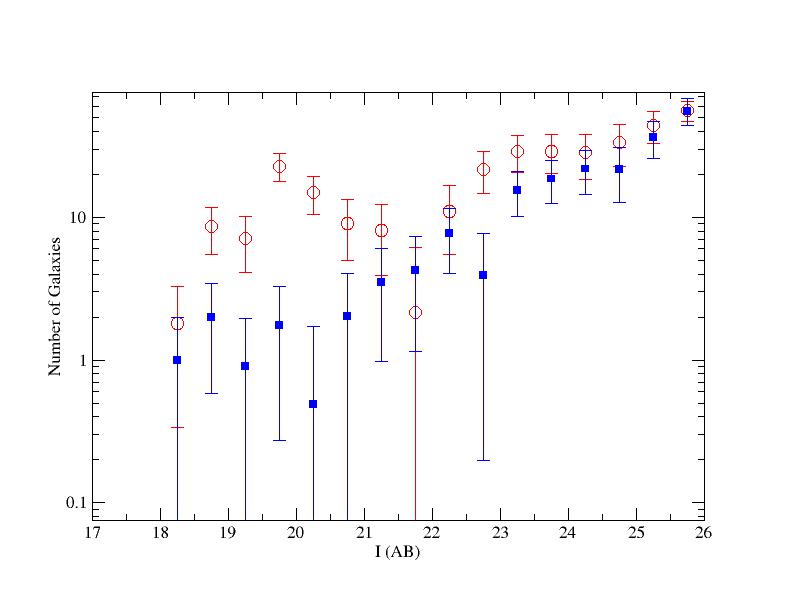} \\
\end{tabular}
\caption{Luminosity functions for red sequence (red circles) and blue cloud (blue squares) galaxies in 
the Eastern (left) and Western (right) subclusters of the Bullet Cluster.}
\label{bulletrbld}
\end{figure*}

Figure~\ref{bulletrbld} shows the derived red sequence and blue cloud luminosity functions for the two
subclusters in the Bullet. The shape of the red sequence luminosity function is not well represented by a 
single monotonic Schechter function, but shows a 'dip' or 'plateau' at intermediate luminosities ($-18 < M_I 
< -20$), followed by a power-law rise at lower luminosities. These may be best fitted by a double Schechter 
function or a single Schechter function plus a power-law as in \cite{Phillipps1995} and \cite{Popesso2006}. 
The strength of this dip appears to depend on the richness of the cluster being observed. There is a clear lack 
of intermediate luminosity galaxies in the poorer Western subcluster whereas the dip shows as a flattening of 
the cluster number counts in the richer Eastern object, which is not easily distinguishable (at least in a
statistical sense)  from a single Schechter function. Similar behaviors are observed in other clusters in the
sample as well: there is a clear inflection in the luminosity functions of red sequence galaxies in Abell 520 
for example and in Abell 2744, but this is less evident in other clusters. The red sequence luminosity function
of relatively massive nearby Abell clusters in \cite{Barkhouse2007} presents a similar behaviour, while
\cite{Popesso2006} find a 'double-Schechter' fit to the total luminosity functions of clusters in their
SDSS/REFLEX sample.

\section{Results}

Here we discuss the luminosity functions, colour-magnitude relations and red sequence/blue cloud luminosity
distributions for our full sample of collisional clusters and a comparison sample of clusters regarded as non-interacting,
with the analysis as carried out above for the Bullet Cluster.

\subsection{Luminosity Functions for Collisional Clusters}

We show the luminosity function parameters $m^*$ (the apparent characteristic luminosity) and
$\alpha$ (the faint-end slope) for the best fits to a single Schechter function for all clusters (and
their subclusters) in Table 2. The best fits and the relative error ellipses are presented in 
a series of figures similar to Figure~\ref{bulletlfs} in the Appendix (we do not show these in the 
printed version in order to save space and focus the attention of the reader on the scientific results 
of this analysis).

\begin{table*}
 \centering
 \begin{minipage}{0.8\textwidth}
  \caption{Luminosity Function parameters}
  \begin{threeparttable}
  \begin{tabular}{lcccc}
  \hline \hline
   Cluster & $m^*_I$ & $\alpha$ & CMR Slope & Intercept\\
   \hline
A520\tnote{a} & $18.19 \pm 0.81$ &$ -0.90 \pm 0.51 $ & $-0.0327$ & 0.872\tnote{e}\\
A520\tnote{b} & $17.71 \pm 0.63$ & $-0.97 \pm 0.46 $ &                     &\\
A520\tnote{c}  & ...                             & ...                              &                      &\\
A520\tnote{d} & ...                              & ...                             &                      & \\
A2163\tnote{a} & $17.90 \pm 0.44$ & $-1.27 \pm 0.04$ & $-0.0315 $ & 0.708\tnote{e} \\
A2163\tnote{b} & $17.78 \pm 0.42 $ & $-1.19 \pm 0.05$ &                    &             \\
A1758\tnote{a} & $19.20 \pm 0.44$ & $-0.71 \pm 0.26$ & $-0.0440$ & 0.886\tnote{f}\\
A1758\tnote{b} & $18.72 \pm 0.34 $ & $-1.07 \pm 0.10$ &                 &             \\
Bullet Eastern & $18.25 \pm 0.49 $ & $-1.27 \pm 0.05$ & $-0.0393$ & 0.865\tnote{f} \\
Bullet Western & $18.37 \pm 0.62 $ & $-1.25 \pm 0.07 $ &                   &            \\
Abell 2744 & $18.88 \pm 0.21$ & $-0.95 \pm 0.08$ & $-0.0366$ & 0.895\tnote{f} \\
MACS0553.4-3342 & $20.37 \pm 0.25$ & $-0.78 \pm 0.12$ & $-0.0356$ & 1.070\tnote{g} \\
MACS1226.8+2153 Center & $21.03 \pm 0.48$ & $-0.03 \pm 0.83$ & $-0.0556$ & 1.065\tnote{g} \\
MACS1226.8+2153 NE & $20.19 \pm 0.48$ & $-0.76 \pm 0.21$ &  & \\
MACS1226.8+2153 S & $20.23 \pm 0.26$ & $-0.91 \pm 0.14$ & & \\
MACS0358.8-2955 & $19.09 \pm 0.26$ & $-1.06 \pm 0.08$ & $-0.0619$ & 1.045\tnote{g} \\
   J0916+2951 South & $20.35 \pm 0.44$ & $-1.22 \pm 0.19$ & $-0.0762$ & 1.257\tnote{g} \\
   J0916+2951 West & $20.80 \pm 0.58$ & $-0.85 \pm 0.24$ & & \\
   J0717+3745 & $20.56 \pm  0.16$ & $-0.80 \pm 0.08$ & $-0.0385$ & 1.219\tnote{g} \\
   CL0025-1222 & $20.15 \pm 1.02$ & $-1.02 \pm 0.34$ &  & \\
 \hline
\end{tabular}
     \begin{tablenotes}
       \item[a] Position 1
       \item[b] Position 2
       \item[c] Position 3
       \item[d] Position 4
       \item[e] Fit to all positions, intercept for $I=18.0$
       \item[f] Fit to all positions, intercept for $I=19.0$
       \item[g] Fit to all positions, intercept for $I=20.5$
     \end{tablenotes}
  \end{threeparttable}
\end{minipage}
\label{lfpar}
\vspace{1.0cm}
\end{table*}

Within errors, $M^*$ and $\alpha$ are very similar for clusters in the same redshift ranges, although 
errors on $M^*$ are above 0.3 mag. (this is poorly determined for single clusters because of small 
number statistics and the difficulty in fitting such a steeply varying function at the bright end) while typical 
errors on the faint-end slope $\alpha$ are of the order of 20\%. We find $M^*_I \sim -22$ (including
an $e+k$-correction from a \citealt{Bruzual2003} model assuming a formation redshift $z_f = 3$, a 
1~Gy e-folding time and solar metallicity),  and $\alpha \sim -1$, over a range of up to 8 magnitudes 
(a factor of 1,000 in luminosity) reaching well into the regime of true dwarf galaxies, with $M_I \sim -15$, 
about 600 times fainter than the Milky Way and resembling Local Group dwarf spheroidals such as the 
Fornax and Carina dwarfs. 

We derive a composite luminosity function for clusters within small redshift intervals, in our case at $<z> 
\sim 0.25$, $\sim 0.42$ and $\sim 0.55$, to improve our estimates of $M^*$ and $\alpha$, using the
method by \cite{Colless1989}: this assumes that there is relatively little variation from cluster to cluster, 
which is consistent with the individual luminosity functions shown in Table 2.

\begin{figure*}
\includegraphics[width=\textwidth]{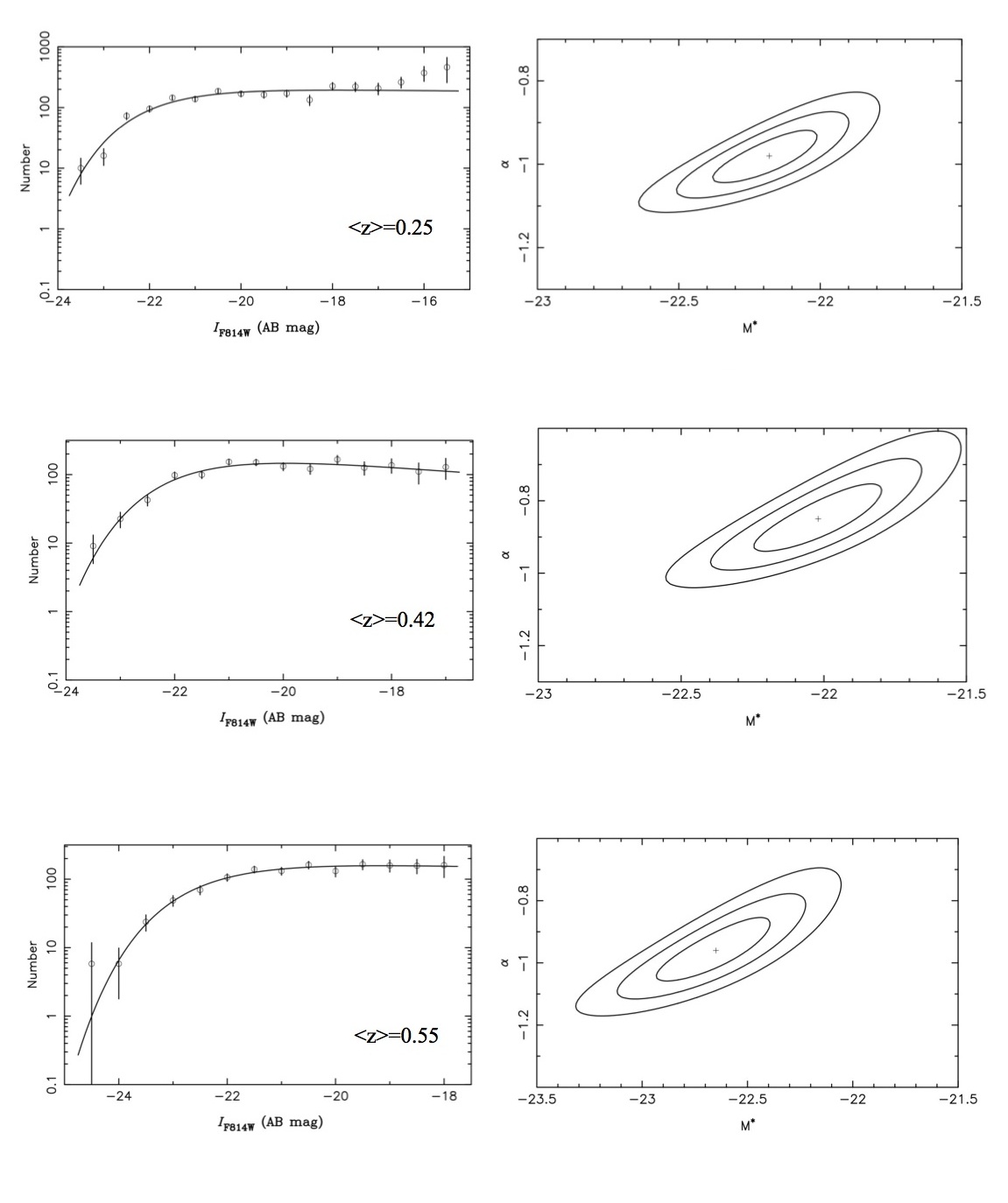}
\caption{Composite luminosity functions and best  Schechter fit, with error ellipses to the side
for clusters at $<z>=0.25$, 0.42 and 0.55 as identified in the caption. The best values for the
parameters and their conditional $1\sigma$ errors are shown in Table~\ref{clus}.}
\label{zclf}
\end{figure*}

Figure~\ref{zclf} shows the composite luminosity function for all collisional clusters and the 
best fitting single Schechter function with the corresponding error ellipse. We have again used the 
model by \cite{Bruzual2003} to shift all the data (for distance modulus and $k$-correction) to 
$z=0$. 

We find $M^*_I=-22.20 \pm 0.12$ and $\alpha=-0.98 \pm 0.04$ for the five clusters with $0.20
< z < 0.31$. For the three collisional clusters at $0.41 < z < 0.43$; the best fitting parameters 
are: $M^*_I=-22.04 \pm 0.13$, $\alpha=-0.86 \pm 0.07$. The composite luminosity function
for the three clusters at $0.53 < z < 0.58$ has  best fitting parameters of: $M^*_I=-22.68
\pm 0.13$ and $\alpha=-0.97 \pm 0.07$. 

\subsection{The colour-magnitude relations}

\begin{figure*}
\includegraphics[width=\textwidth]{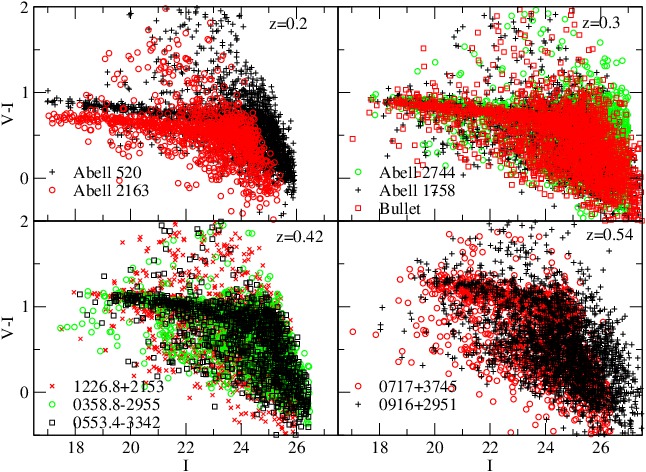}
\caption{Colour-magnitude relations for galaxies in the clusters examined in this paper. We show
all clusters in each redshift range in the same figure. The clusters and redshift ranges are indicated
in the figure legends. We omit CL0025-1222 as its $V_{F555W}$ colour is very different from that of
all other objects. With the exception of Abell 2163, where the extinction coefficients are more 
uncertain, we find that all clusters are consistent with a single  colour-magnitude relation. The slope 
and intercept of the relation are also consistent with passively evolving stellar populations formed 
at high redshift.}
\label{compcmds}
\vspace{1cm}
\end{figure*}

Figure~\ref{compcmds} plots the colour-magnitude diagrams for all galaxies in the collisional
clusters. We show the two clusters at $z=0.2$, the three at $z \sim 0.3$ and then those at
$z \sim 0.42$ and at $\sim 0.55$, each in the same diagram. Figures for all individual clusters
are shown in the Appendix. The best fitting slopes and intercepts to a straight line (using a robust
fitting method) are given in Table 2. In order to better show the cosmic variation in colour
for massive cluster galaxies we calculate the intercept at the apparent magnitudes cited in Table 2,
which is chosen to be close to the $M^*$ point. All clusters contain well-defined red sequences, which 
can be fitted by a single straight line, and can be followed for nearly 7 magnitudes (and sometimes more) 
to the photometric limits of the data The red sequences appear to be very narrow, with a scatter dominated 
by photometric errors. Even collisional clusters are therefore largely composed of a population of quiescent 
galaxies, similar to local objects. Comparison (non-collisional) clusters (Figures in the Appendix) also show 
the same behaviour, with tight red sequences having colours consistent with those of the main sample of 
`bullet-like' objects.

The red sequences essentially overlap for all clusters at the same redshift: red galaxies
have the same colour irrespective of environment in all clusters, even in different stages
of a collision (or no collision). In A2163 there is evidence that the extinction may be different 
than stated in \cite{Schlafly2011}; this is not surprising as the $A_V$ is close to 1 mag. and 
foreground extinction may be patchy. CL0025-1222 has a different $V$ band and this yields a 
substantially different colour, so this is not included here. With these exceptions, the colour-magnitude 
relations are remarkably consistent, including for non-collisional clusters. In addition, the red sequence 
colours at different epochs are also consistent with our adopted passive evolution model. 

\subsection{Red Sequence and Blue Cloud Luminosity Functions}

\begin{figure*}
\includegraphics[width=\textwidth]{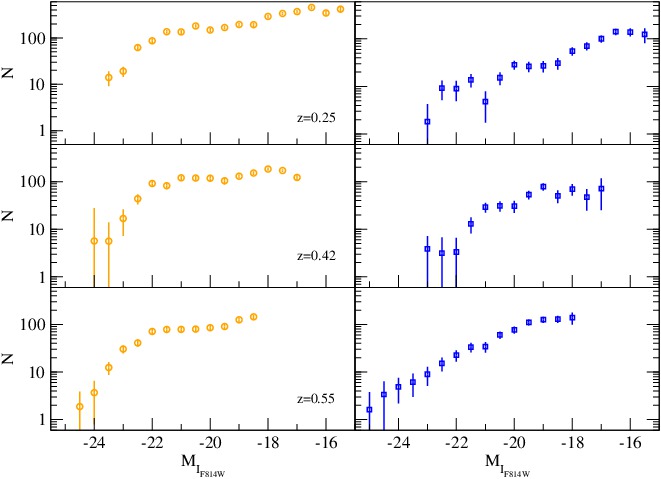}\\
\caption{Composite luminosity distributions for red sequence (orange circles) and blue cloud galaxies 
( filled blue squares) in the collisional clusters examined in this paper. From top to bottom the mean redshifts
are $z=0.25$, $0.42$ and $0.55$. The red sequence luminosity distributions resemble a Schechter
function with an extra power-law or double Schechter component at the faint end. There is no evidence
that the faint end weakens in this sample with redshift. The blue cloud galaxies follow a truncated power
law: they are rare at $z=0.25$ and brighter than the 'dip' in the red sequence, but they become slowly
more important at higher redshifts and bright blue galaxies are observed in the highest redshift bin.}
\label{complfs}
\end{figure*}

We derive composite luminosity functions for red sequence and blue cloud for collisional
clusters at $<z>=0.25$ (A520, A1758, A2163, A2744 and the Bullet), $0.42$ (MACS0553.4-3342,
MACS1226+2153 and MACS0358.8-2955) and $0.55$ (J0916+2951 and J0717+3745, CL0025-1222
has a very different $V$ band, although its $I$ data are included in the total luminosity functions in
Figure~\ref{zclf}) in Figure~\ref{complfs})).

We fit a single Schechter function to the red sequences, in order to better parametrise their evolution.
For galaxies at $<z>$ =0.25, $0.42$ and $0.55$ we find, respectively, $M^*_I=-22.72 \pm 0.13$, 
$-22.46 \pm 0.23$, $-22.98 \pm 0.21$, and $\alpha=-1.20 \pm 0.03$, $-1.04 \pm 0.03$, $-1.12 \pm
0.06$, where these have been corrected for distance modulus and $k$-correction to $z=0$, but not for the
$e$-correction to better show the evolution over this redshift range. The above values are consistent with the
pure passive evolution of our simple model with $z_f=3$ and $\tau=1$ Gyr. By appling the appropriate
shifts for distance modulus and the $k+e$ corrections, the red sequence luminosity functions effectively
overlap. Red sequence galaxies therefore appear to evolve passively and  the red sequence appears to be 
well established, to at least $M_I \sim -18$ (5 magnitudes below the red  galaxy $M^*$) in clusters out to 
at least $z=0.6$.

However, the red sequence luminosity distributions are not very well fitted by a single Schechter function. 
They appear to show a dip/plateau at intermediate luminosities followed by a power-law or Schechter-like
rise, as observed more clearly in some of the lower mass objects in our sample (e.g., see above for the
Bullet Cluster). A double Schechter function better fits the red sequence in local clusters such as Coma 
and the nearby objects studied by \cite{Barkhouse2007} as well as the field LF of \cite{Loveday2012}.
The reduced $\chi^2$ is $3.5$ for $z=0.25$, 2.0 for $z=0.42$ and $1.1$ for $z=0.55$ for a single 
Schechter function; the better fit at high redshift is due to the fact that we are losing statistical power 
at the faint end. 

We attempt to fit a double Schechter function to the $z=0.25$ data: this yields $M^*_{bright}=-22.72 \pm 0.13$, 
$\alpha_{bright}=-1.20 \pm 0.03$, $M^*_{faint}=-14.81 \pm 1.82$ and $\alpha_{faint}=-0.13 \pm 3.06$. Unfortunately,
there is little statistical power to fit the faint Schechter function, even if we fix the $\alpha_{bright}$ value to $-1$
as done in \cite{Barkhouse2007}. Nevertheless, inspection of Figure~\ref{complfs} for the red sequence (left-hand
panels) galaxies, shows a dip/plateau at intermediate magnitudes, as seen in other local clusters and in the
compilation of \cite{Barkhouse2007}, as well as in the field red sequence luminosity function by \cite{Loveday2012}.

For blue cloud galaxies we observe that, in the lowest redshift bin of Figure~\ref{complfs} there are few
bright blue galaxies brighter than the 'dip' in the red sequence at $M_I \sim -20$, while the fainter blue
galaxies approximately follow a power-law of slope $\sim -1.4$. In the two higher redshift bins we see
both a relative increase in the fraction of blue galaxies, especially at the faint end, and the presence of
some bright blue galaxies which are absent in local samples. This may be explained by either a decreasing
quenching efficiency in higher redshift clusters (although the environment in these collisional objects should
be hostile to star-forming galaxies, as observed by \citealt{Ma2010} in CL0025-1222) or it may reflect
(as it does for local clusters -- \citealt{DePropris2003b}) the luminosity function for blue galaxies in the
field, where such objects are 'interlopers' in the cluster fields and are quickly quenched upon infall.

\begin{table*}
 \centering
 \begin{center}
 \begin{minipage}{0.8\textwidth}
  \caption{Non-collisional clusters studied in this paper}
  \begin{threeparttable}
  \begin{tabular}{lcccccc}
  \hline \hline
   Cluster & $z$ & Passband & Exposure &  Area  & Proposal & PI \\
               &        &                &         [s]            &       [arcmin$^2$]\\
 \hline
MACS0547-3904 & 0.21 & $V$ & 1200 & 11.87 & 12166 & Ebeling \\
                                 &          & $I$ &   1440 & 11.87 &             &                \\
Abell 1351 & 0.32 & $V$ & 1200 & 11.87 & 12166 & Ebeling \\ 
                    &           & $I$ & 1440 &   11.87 &             &                \\
MACS0417.5-1154\tnote{a} & 0.44 &$V$ & 1788 & 10.62 & 12009 & Von der Linden  \\
                                    &           & $I$ & 1910 & 12.23 &                              &             \\                        
MACS1621.3+3810\tnote{b} & 0.47 & $V$ & 1200 & 11.87 & 12166 & Ebeling \\
                                     &          & $I$  & 1440 & 11.87 & & \\  
CL0016+16 & 0.54 & $V$ & 2240 & 39.17\tnote{c} & 10635 & Ziegler \\
                       &          & $I$ & 4560 & 11.71 & 11560 & Ebeling \\
 \hline

\end{tabular}
     \begin{tablenotes}
       \item[a] Combined $V$ and $I$ coverage is 6.79 arcmin$^2$
       \item[b] Combined $V$ and $I$ coverage is 9.28 arcmin$^2$
       \item[c] $2 \times 2$ mosaic
       \end{tablenotes}
  \end{threeparttable}
\end{minipage}
\end{center}
\label{compdat}
\end{table*}

\subsection{Comparison with other clusters}

We have studied a small sample of comparison clusters, as part of a larger dataset, to understand 
whether collisional clusters exhibit a different behaviour and isolate the effects of the mergers on the
cluster members. Table 3 shows the same information as in Table 1 for the
non-collisional objects (cluster identification, redshift, bands, exposure times, proposal ID and 
PI).

We treat these objects in the same way as the collisional clusters. Their luminosity functions,
colour-magnitude relations and luminosity distributions for red and blue galaxies can be found
in the appropriate figures in the Appendix. Table 4 shows the derived parameters for
a single Schechter function fit to all galaxies. Table 4 also presents the slope and intercept
of the robust fits to the colour-magnitude relations (cf. Table 2 for the collisional clusters).

\begin{table*}
\centering
 \begin{minipage}{0.75\textwidth}
  \caption{Luminosity Function and colour-magnitude parameters for non-collisional clusters}
  \begin{threeparttable}
  \begin{tabular}{lcccc}
  \hline \hline
   Cluster & $M^*_I$ & $\alpha$ & Slope & Intercept \\
   \hline
MACS0547-3904 & $19.97 \pm 0.77$ & $-0.89 \pm 0.37$ & $-0.0598$ & $0.961$\tnote{a} \\
Abell 1351             & $18.73 \pm 0.47$ & $-1.01 \pm 0.10$ & $-0.0436$ & $0.946$\tnote{a} \\
MACS0417.5-1154 & $20.28 \pm 0.29$ & $-0.88 \pm 0.12$ & $-0.0794$ & 1.552\tnote{b} \\
MACS1621.3+3810 & $20.23 \pm 0.64$ & $-0.91 \pm 0.36$ & $-0.0584$ & 1.184\tnote{b} \\
CL0016+16 & $20.97 \pm 0.44$ & $-0.65 \pm 0.31$ & $-0.0639$ & 1.318\tnote{c} \\
 \hline
\end{tabular}
     \begin{tablenotes}
       \item[a] Intercept at $I=19.0$
       \item[b] Intercept at $I=20.0$
       \item[c] Intercept at $I=20.5$
       \end{tablenotes}
 \end{threeparttable}
\end{minipage}
\label{lfpar2}
\vspace{1cm}
\end{table*}

In general, we recover the same pattern as observed for collisional clusters. A passively evolving
$M^*$ coupled with a non-evolving $\alpha$ (despite the large errors for each individual cluster); 
tight and well-defined colour-magnitude relations with similar colours to the collisional clusters, 
and which can be followed for several magnitudes beyond the $M^*$ point and are consistent with passive 
evolution between the observed redshifts. Luminosity distributions for red sequence and blue cloud galaxies 
broadly follow the same scheme: a double Schechter function for red galaxies, with a dip or plateau at 
intermediate magnitudes, while blue galaxies are rare brighter than this feature but become progressively 
more important with redshift. However, as discussed above, the precise features of the dip are difficult
to establish in a general sense because of small number statistics.

This implies that what we are observing in collisional clusters can broadly be extended to the entire
population of galaxies in clusters and that the effects of environment have been both weak and very
similar, irrespective of the cluster properties and its dynamical history, to a redshift of at least $z=0.6$,
although of course environmental effects may have been important at earlier lookback times. It is likely 
that cluster galaxies have largely completed their evolution at least before the collision took place; in this
fashion, even the violent cluster environment in a merger is essentially unable to affect the properties of 
its member galaxies (via ram stripping).

\section{Discussion}

We have derived deep luminosity functions for all galaxies in clusters of galaxies at $0.2 < z < 0.6$, both in
objects that appear to be undergoing major and complex collisions (e.g., the Bullet cluster) and for a smaller 
reference sample of seemingly normal objects. We used archival data from the HST (usually taken to study
weak lensing and reconstruct cluster masses) to determine the colour-magnitude relation and derive the 
luminosity distributions for galaxies on the red sequences and blue cloud. Cluster members were identified
statistically, via photometry in reference fields believed to represent the general galaxy population, chiefly
the CANDELS and GOODS surveys. By using clusters in various stages of the merger process (from objects
such as Abell 520 where the cluster appears to be coalescing from at least three separate groups, to the
Bullet cluster where the two clusters have crossed each other, to more complex systems such as Abell 2744
where a very intricate merger between multiple components is taking place) we hope to isolate the effects
of environment, and specifically to separate dark matter halos from the cluster gas. 

\subsection{Impact of mergers on galaxy luminosities and colours}

\begin{figure}
\includegraphics[width=0.475\textwidth]{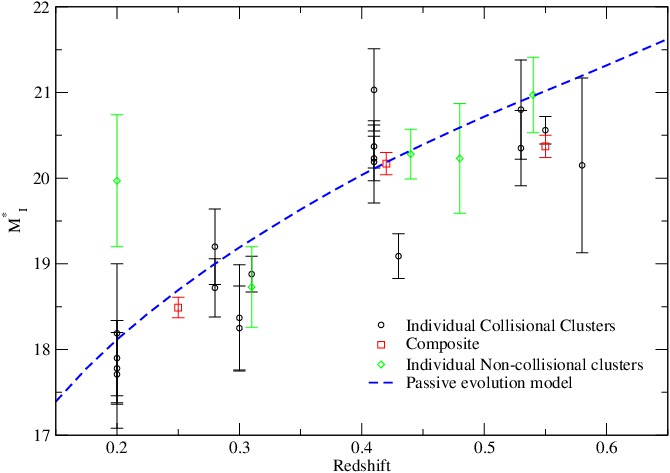}
\caption{Evolution of the $M^*$ point with redshift for collisional clusters, the composite luminosity
functions we show above, and the comparison non-interacting objects. The blue dashed line shows
a passive evolution model as in the text, with $M^*_I=-22$ at $z=0$. The various categories of clusters 
are identified in the figure legend. This shows the essentially passive evolution of the characteristic 
luminosity of galaxies, irrespective of cluster properties or dynamical status}
\label{mstarevol}
\vspace{0.5cm}
\end{figure}

In all our clusters, irrespective of their dynamical status, whether `collisional' or `normal' objects, we observe
that  $M^*$ evolves in a manner consistent with the passive stellar evolution of high metallicity populations 
formed at high redshift in short star formation episodes, to at least $z=0.6$. We plot the results for all clusters
and the composite luminosity functions (in Figure~\ref{zclf}) vs. redshift in Figure~\ref{mstarevol} where 
we also show the predicted evolution for a pure passive model normalized to $M_I=-22$ at $z=0$. This is
consistent with our earlier work in the infrared \citep{DePropris1999, DePropris2007}, as well as optical
observations (e.g., \citealt{Andreon2008}). Massive cluster galaxies therefore appear to have assembled 
their stellar masses before the redshift of observation; current limits to this may be as high as $z \sim 1.5$
and possibly beyond \citep{Mancone2010}. 

\begin{figure}
\includegraphics[width=0.475\textwidth]{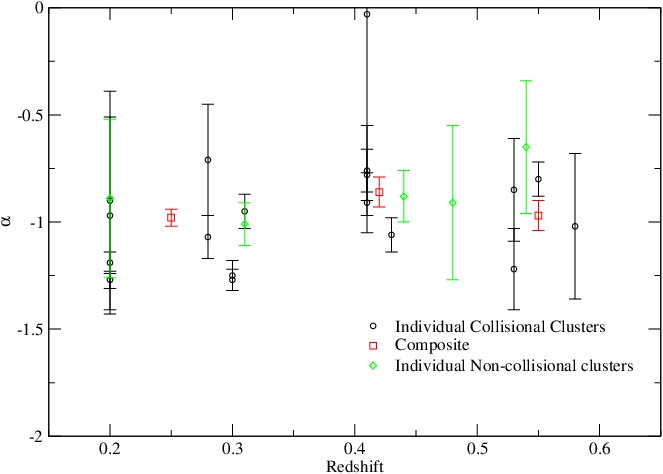}
\caption{Evolution of the faint end slope $\alpha$ with redshift for all clusters we consider in this
paper. Symbols are identified in the figure legend. We see no evidence that $\alpha$ differs significantly
from $-1$ over the redshift range we sample}
\label{alphaevol}
\end{figure}

We also find that the faint end slope $\alpha$ is $\sim -1$ in all clusters and at all redshifts, irrespective of
cluster properties, as shown in Figure~\ref{alphaevol}. The dwarf galaxy luminosity function therefore does not
appear to evolve significantly, at least down to $M_I \sim -16$, out to $z=0.3$, and at least to $M_I=-18$ 
(4.5 magnitudes below $M^*$), out to $z=0.6$. This is true for all collisional clusters  as well as the
comparison  clusters. This is consistent with the findings of \cite{Andreon2008} and the more recent work 
by \cite{Mancone2012}, as well as our composite luminosity functions of massive  clusters at $<z>$ =0.25 
in \cite{Harsono2009}. 

\subsection{Passive evolution of the red sequence}

All clusters we study are also dominated by red sequence galaxies.  We observe very well formed and tight 
colour-magnitude relations that extend to well below the $M^*$ point -- up to 7 magnitudes in some clusters. 
This is similar to what is observed in local clusters such as Coma \citep{Eisenhardt2007, Hammer2010} 
where the  red sequence is traced to $M_I \simeq -13$,  and  Abell 1185 \citep{Andreon2006b},  among 
others. The  colour-magnitude relations are very similar from cluster to  cluster, irrespective of whether 
the cluster is involved in a collision or otherwise; this is true even for subclusters  within, e.g., Abell 1758, 
Abell 2163, J0916+2951 or the Bullet Cluster, and even for the multiple groups (and the intragroup region) 
of Abell 520 where the cluster has not yet coalesced from the individual components (e.g.,
Figure~\ref{compcmds}).  

Environmental effects on the red sequence galaxies must therefore have been relatively minor. The
cluster collisions, which are observed here in different stages (from initial interactions like in Abell
520 to about 0.7 Gyr after the event in J0916+2951, including complex on-going mergers such as
Abell 2744), do not appear to have affected the properties (luminosities and colours) of red sequence
galaxies. In collisional clusters galaxies have been subjected to an abnormally strong ram stripping
wind, at least two orders of magnitude more powerful than they would otherwise encounter in the cluster
environment. The observation that the cluster collision has not affected the red galaxies can be explained 
only if these objects were completely quiescent (gas free) before the interaction took place and implies 
that the stellar populations of these galaxies were in place at least since before the two (or more) clusters 
merged. For J0916+2951 \cite{Dawson2012} claim that the galaxies are observed 0.7 Gyr after the crossing 
time. This sets the epoch of (red) dwarf galaxy formation to at least $z=0.7$ or about 7 Gyr ago.

\subsection{Evolution of red cluster dwarfs}

From our data we can select galaxies on the red sequence and the blue cloud (see Figures~\ref{coldist}
and ~\ref{colprob}) and determine luminosity functions for these objects to measure their differential
evolution. Composite luminosity distributions for galaxies at $<z>$ =0.25, 0.42 and 0.55 are plotted
in Figure~\ref{complfs}. We see no evidence that either $M^*$ or $\alpha$ for red sequence galaxies
 evolve in any significant way, except the pure luminosity evolution implied by the ageing of stellar 
populations formed at high redshift. This implies that red sequence galaxies down to $M_I \sim -18$
were already formed at least by $z=0.6$ (cf. \citealt{Andreon2008} for galaxies to $M^*+3.5$ in the rich 
$z=0.83$ cluster MS1054-03). As we observe a similar behavior for both the sample of eleven collisional 
clusters and the five non-interacting objects at similar redshifts, this suggests that environmental
influences on red sequence galaxies have been weak and that red cluster dwarfs have been quiescent 
since long before the cluster mergers.

However,we also note that the red sequence luminosity function is not well fitted by a single Schechter
function and is best represented by a double Schechter function, with a dip or plateau at intermediate
magnitudes and a subsequent rise. This is most evident in lower mass clusters like A520 and the smaller
Bullet subcluster. Similarly shaped distributions for red sequence galaxies in local clusters have been 
found by \cite{Barkhouse2007} among others. We therefore witness no change in the luminosity distribution
of red galaxies, other than the apparent brightening due to passive evolution.

While this appears to be in contrast with the observations by \cite{Smith2012} on the age distribution
of galaxies in the Coma cluster, it must be noted that these objects are brighter than the faint Coma
dwarfs where \cite{Smith2012} find evidence of younger ages; additionally, the young dwarfs tend to
lie in the Coma outskirts, while those in the core are uniformly old. The cluster regions we sample 
here tend to lie within $\sim 500$ kpc or less, and are therefore more representative of high density
regions, so there is not necessarily a disagreement between our work and age determinations (based
on spectroscopy) for dwarfs in Coma and elsewhere.

The evolution of the faint end of the luminosity function for dwarf galaxies has been the subject of numerous 
papers. \cite{DeLucia2007} have used a sample of clusters at $0.4 < z < 0.8$ to claim that the  red sequence
dwarf-to-giant ratio (measured as the ratio of galaxies between two luminosity intervals) weakens with 
redshift at $z > 0.4$ at least. \cite{Rudnick2009} confirm this result based on a more complex re-analysis 
of the EDisCS data. A weakening of the faint end of red sequence in more distant clusters is also claimed by
\cite{Stott2007,Lerchster2011,Lemaux2012} and by \cite{Rudnick2012}. However, \cite{Crawford2009} observe 
no weakening of the red sequence in their sample, while \cite{Andreon2008} also does not observe any
significant evolution of the faint-end slope $\alpha$ out to high redshift. We believe that our analysis, 
which benefits from the use of HST data and extensive comparison fields for statistical identification of 
cluster galaxies (taken under the same conditions), confirms the counter-claims by  \cite{Andreon2008} 
and \cite{Crawford2009} as to the (lack of) evolution of red sequence galaxies in clusters.

One obvious caveat is that we may be comparing objects of different masses and therefore in different 
stages of dynamical evolution. In general, the clusters studied by the EDisCS group tend to be relatively 
less massive objects than the richer systems analysed by \cite{Harsono2007,Harsono2009} and some of 
our targets. If galaxy evolution depends, as one would expect, on the mass of the parent halos (e.g.
\citealt{Grutz2012}), dwarf galaxies may be particularly affected by the different environments. On the other 
hand, we observe no environmental dependence even when we look at clusters with different masses and
evolutionary histories, and we see no strong environmental effect when we study the lower mass subclusters 
in some of our targets (e.g., in Abell  520,  A1758, A2163, J0916+2951 and the  two Bullet subclusters, that 
can be well distinguished -- see Table~4 and Figures~\ref{mstarevol} and ~\ref{alphaevol} above).

\subsection{Evolution of blue cluster galaxies}

We now focus on the blue galaxies. These appear to be well fitted by a power-law or single Schechter function. 
However, we observe significant evolution in these objects. At low redshifts, there are few, if any, luminous blue 
galaxies in clusters, especially brighter than the `dip' in the red sequence luminosity function. The luminosity
distribution fainter than $M_I \sim -20$ may be fitted by a power law of slope $\sim -1.4$. However, at higher
redshifts we see both an increase in the fraction of blue galaxies, especially at the faint end, and an increasing
contribution from more luminous blue galaxies. \cite{Raichoor2012} argue that the blue fraction (although 
our objects are more representative of the blue cloud population rather than the classical blue fraction) 
evolves according to redshift, luminosity  and environment (see also \citealt{DePropris2003a}). Similarly,
in Abell 868 \cite{Boyce2001} and \cite{Driver2003} show that late-type galaxies come to dominate the
luminosity function at the faint  end, with a slope of $\sim -1.4$.This is consistent with a model where 
such galaxies are quenched to join the red sequence, probably after  fading \citep{DePropris2003b,Peng2010}. 

Because this does not appear to alter the red sequence luminosity function or the total luminosity function
(see above) the evolution must be largely in density, with little contribution from mergers. The true masses 
of these objects are likely to be lower and their apparent luminosities are boosted by star formation
\citep{DePropris2003a,Holden2009}. This is consistent with a `downsizing' model for blue galaxies in 
clusters. 

There are two possible ways in which this might occur. On the one hand, quenching may become more 
efficientin local clusters, as these become more massive. The observations by \cite{Lemaux2012} support 
this hypothesis, as more massive blue cloud galaxies in the CL1604 supercluster appear to be quenched 
earlier in the more dynamically relaxed systems. \cite{Tajiri2001} present a model where blue galaxies may
continue forming stars within clusters until they are ram-stripped, a phenomenon that would become more
prominent as the cluster relaxes. However, here we observe similar behaviours across a variety 
of clusters, including objects which are by no means relaxed and are involved in major mergers and more
quiescent systems.

The second possibility is that blue galaxies are a transient population, quickly quenched upon infall.
Their luminosity function should therefore reflect the luminosity function of field blue galaxies. At higher
redshifts we expect that more massive and luminous blue galaxies will be present in the field (because of
downsizing). This would account for the growth in the blue fraction with redshift  and its dependence on 
galaxy luminosity and distance from the cluster centres \citep{Raichoor2012}. \cite{Haines2009}  reach a 
similar conclusion where the blue population is drawn from the field but interactions boost the star formation
signal at large cluster-centric radii. \cite{Li2012} study blue cloud galaxies within the  Red Sequence Clusters
Survey 2 with spectroscopy from extra fibers in the WiggleZ dataset and they find similar results  to ours, but 
are also unable to resolve the discrepancy between an increasing quenching efficiency and a form of rapid 
quenching in clusters, where the blue population is drawn from the general field.  This can also be inferred 
from the analysis of \cite{Ma2010} in Cl0025-1222 where the starbursts are not related to the collision but are
observed primarily in infalling galaxies.

\begin{figure}
\hspace{-0.75cm}
\includegraphics[width=0.4\textwidth, angle=-90]{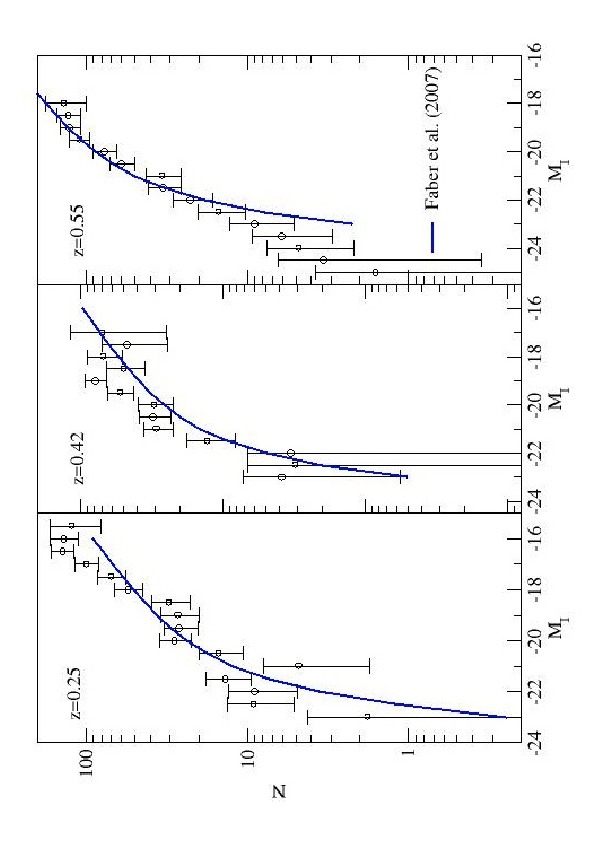} 
\caption{Luminosity distributions for blue sequence galaxies in clusters, together with the evolving
field blue galaxy luminosity function from Faber et al. (2007). The data are consistent with cluster blue
galaxies being drawn from the field and having the same properties.}
\label{bluefab}
\end{figure}

Figure~\ref{bluefab} compares the blue cloud luminosity functions for galaxies in clusters with those derived
by \cite{Faber2007}, assuming $B-I=1$ and with arbitrary normalization; since \cite{Faber2007} does not
quote an $\alpha$ we hold this to the local value of $-1.3$. We see that the field blue luminosity function
(although the definitions of blue galaxies are somewhat different) is generally consistent with the cluster blue 
luminosity function, with the only possible exception of an excess of bright blue cluster galaxies in the highest
redshift bin, where our statistics are poorer. However, some of these objects have now been detected in the
field, at similar redshifts, by the PRIMUS \citep{Moustakas2013} and VIPERS \citep{Davidzon2013} surveys. 
This is consistent with the above picture of infall and rapid quenching.

In local clusters at $z < 0.1$ (e.g, \citealt{DePropris2003b}), the `star-forming' galaxy luminosity function 
appears to be identical to the field one, while the field 'red sequence' galaxy luminosity function is lacking in 
faint objects, which are instead present in the clusters. It is also interesting to notice that in this case, simply
quenching the `blue' galaxies and adding them to the `red' galaxies in the field yields a good match to the 
luminosity function of red cluster galaxies (although these are selected spectroscopically). However, it is not 
clear that this mechanism can also explain the observed evolution in higher redshift clusters. 

In some ways, our picture is similar to that in \cite{Gilbank2008} from analysis of clusters in the Red
Sequence Cluster Survey 1, where they find an increase in the faint blue galaxy fraction, with brighter
blue galaxies being present at higher redshifts, down to $M_V \sim -19.7$, with a decrease in the
fraction of red sequence galaxies, and little merging, although we do not see the latter effect in our 
data, suggesting that the blue galaxies undergo considerable fading into the faint dwarfs regime at
low redshifts. These objects may provide the 'young' dwarfs observed by \cite{Smith2012} in the Coma
cluster and elsewhere.

\subsection{Implications for Galaxy Formation}

Stellar luminosity and mass functions can be used to constrain the degree of merging that has taken place, 
via its influence on the parameters that control the shape of the luminosity function (e.g., 
\citealt{Drory2008}). 

In the models of \cite{DeLuciaBlaizot2007} about half of the stellar mass is assembled since $z=1$ (albeit 
for brightest cluster galaxies). The tight colour-magnitude relations observed in clusters set strong limits to the 
fraction of wet (gas-rich, star-forming) mergers that may take place. Most of the merging must therefore take
place between gas-poor, spheroidal-like galaxies (which dominate the cluster population).
 
The degree of evolution in $M^*$ that we observe in our cluster galaxies is consistent with pure passive evolution 
of their stellar populations.  If we use the models by Skelton, Bell \& Somerville (2012) we find that there is no room 
for any mass growth above the luminosity increase predicted for a $z_f=3$ model. A model where galaxies undergo 
at least one dry merger between $z=0.6$ and today results in galaxies 0.7 mag. too bright at $z=0$.  Our data constrain 
the mass increase of galaxies to be considerably lower and consistent with none: at least for luminous galaxies mergers 
appear  to be unimportant in the mass assembly history of galaxies since $z < 0.6$.

Dwarf galaxies to $M_I \sim -18$ also appear to be completely assembled at least by this redshift. As 
this is found in all environments we have studied, it appears that the evolution of dwarf galaxies has not 
been strongly affected by the cluster collisions. This is not surprising for red galaxies, where star formation
has likely ceased long in the past and which therefore contain no gas for ram stripping to act on. However,
we also find that the blue cloud objects are also consistent with relatively weak environmental effects,
suggesting that these galaxies are 'recently' accreted from the field (cf. \citealt{Ma2010}).

We use the simulations by \cite{Rudnick2012} to quantify the degree of merging that must have taken
place on the red sequence at the faint end. The observation that the faint end slope, both for all galaxies
and for the red sequence, appears not to have evolved to $z=0.6$ is also inconsistent with a significant
contribution from mergers. Even in the 2-merger model of \cite{Rudnick2012}, $\alpha$ evolves from
$\sim -1$ to $\sim -0.6$,  while our data is more consistent with their no-merger model. 

\subsection{Surface brightness selection effects and the evolution of the red sequence}

At face value, our results are in contrast with the expectations of simple hierarchical models as well as with
the result by the EDisCS group \citep{DeLucia2007}, as well as others, for evolution of the faint end of the
red sequence in clusters. There are two possibilities for this: firstly, the dwarf to giant ratio as derived by
\cite{DeLucia2007} and \cite{Stott2007} for example, is actually measured as the ratio of the number of 
galaxies within two luminosity ranges, for 'giants' and 'dwarfs'. This may be affected by the `dip' in the red 
sequence luminosity function. If this evolves faster than the passive evolution found for giants, such a choice 
may `artificially' cause an apparent evolution in the fraction of dwarf galaxies on the red sequence as 
a function of redshift. \cite{Peng2010} suggest that the double Schechter function has two origins: a 
single, more  Gaussian-like, component for massive galaxies, formed from early mergers or dissipative
collapses, and a  steeper Schechter function for fainter galaxies produced by the quenching of star-forming,
lower-luminosity objects. If this is the case, the transitional magnitude between these two behaviours 
(and at which the  luminosity  function presents a dip or plateau) will evolve faster (relatively) than $M^*$ 
for giants which instead is well fitted by a model for passive evolution of stellar populations formed at 
high redshifts (see above). Therefore, if one adopts a series of luminosity cuts to measure a dwarf-to-giant
ratio, as in \cite{DeLucia2007} or \cite{Bildfell2012}, the differential evolution of the transition magnitude 
with  respect  to the giants results in an apparent evolution of the red dwarf galaxy fraction. 

A second possibility for the discrepancy between our findings and claims for accelerated evolution on the
red sequence may lie in surface brightness selection effects. We believe that this is worth discussing in 
some detail. We have shown from Figure~3 how surface brightness selection effects can strongly affect 
the detection of dwarf galaxies. It is clear that in the derivation of galaxy luminosity functions and 
studying their evolution, especially  at the faint end and at high redshifts, such selection in surface 
brightness can play an important role. Galaxies can only be measured if they are detected in the first place, 
and this requires that their central surface  brightness (presumably  the most luminous portion of the 
galaxies) be above a threshold within a small 'detection' aperture, which is usually chosen to be equivalent 
to  the size of the point spread function. HST images have the advantage of a darker sky and a smaller 
and more stable  point spread function and therefore allow us to study dwarf galaxies to faint limits even 
with comparatively short exposures. In addition this stability can be applied to  reference fields to obtain
homogeneous counts to statistically remove the  background and foreground contributions to the galaxy
counts in the clusters' lines of sight.

We have analysed this issue further, via a series of archival exposures of the $z=0.328$ cluster
MS1358.4+6254. Table~\ref{ms1358data} shows the available HST images (in $i$ and $z_{F850LP}$, 
there are  other bandpasses as well, but these are less suitable for our purposes here) for this cluster. 
These   include a series of exposures of length $\sim 2.5$ks and $\sim 5.5$ks; from these we can 
produce stacks of images with exposure lengths of 11ks in $i$ and $z$ and 15ks in $z$. Note here 
that the first exposures (PID 9292) are offset about $1'$ to the North of the others, while the other two
(9717 and  10325) are at a 90$^{\circ}$ angle from each other but  share the same centre.

\begin{table}
 \centering
 \begin{minipage}{0.5\textwidth}
  \caption{Data for MS1358+62}
  \begin{threeparttable}
  \begin{tabular}{cccc}
  \hline \hline
   Program ID & PI & Filter &Exposure\\
   \hline
9292 & Ford & $i$ & 2600 \\
           &          & $z$ & 2740 \\
9717 & Ford & $i$ & 5470 \\
           &          & $z$ & 4065 \\
           &         & $z$ & 5470 \\
10325 & Ford & $i$ & 5482 \\
             &          & $z$ & 5482 \\     
 \hline
\end{tabular}
\label{ms1358data}
  \end{threeparttable}
\end{minipage}
\vspace{1cm}
\end{table}

Figure~\ref{lfsb} shows the effect of surface brightness selection in detail; we plot central
surface brightness vs. total luminosity as in Figure 3, with different exposures colour-coded as in
the figure caption. It is clear that in deeper exposures not only do we see fainter galaxies, but also
galaxies above the luminosity thresholds but lying below the surface brightness limits. 

\begin{figure}
\includegraphics[width=0.475\textwidth]{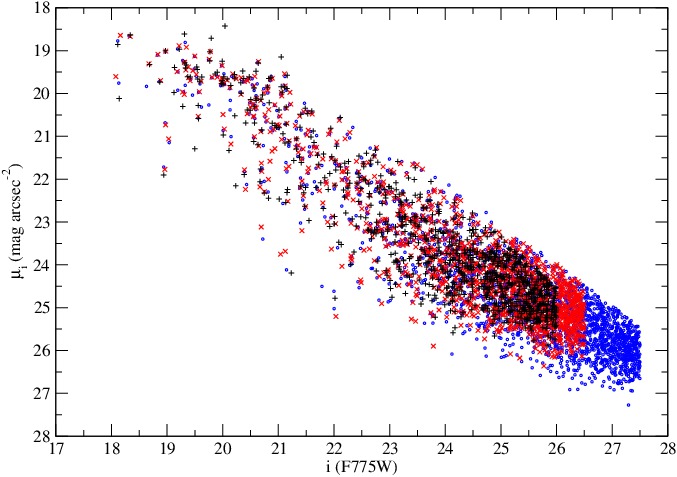} \\
\\
\\
\\
\includegraphics[width=0.475\textwidth]{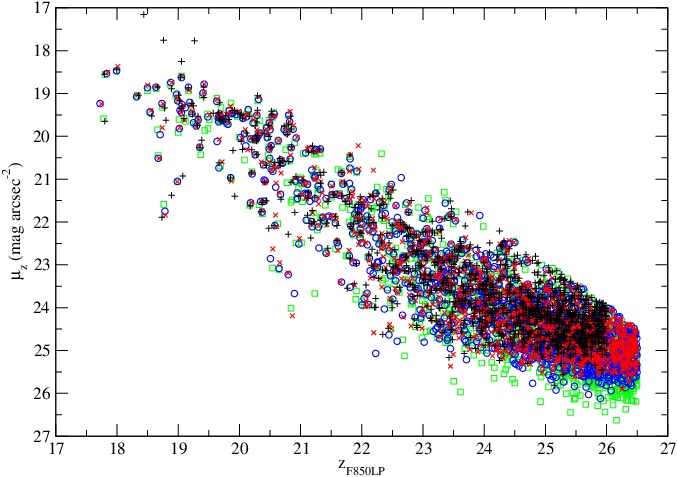} 
\caption{Central surface brightness (used for detection) vs. total magnitude in the $i_{F775W}$
band (Top) and $z_{F850LP}$ (bottom) for galaxies in the MS1358+62 field. The black plus 
signs represent the short 2.6ks exposure (note that this is slightly to the North of the other images), 
while the red crosses show the data for the 5.5ks exposure and the blue circles the 11ks exposure. 
The green squares ($z$ only) are for the 15ks exposure. This shows how increasing depth not only
detects fainter objects but also reveals galaxies above the limiting magnitudes but hidden in the 
night sky}
\vspace{1cm}
\label{m1358sb}
\end{figure}

\begin{figure}
\includegraphics[width=0.475\textwidth]{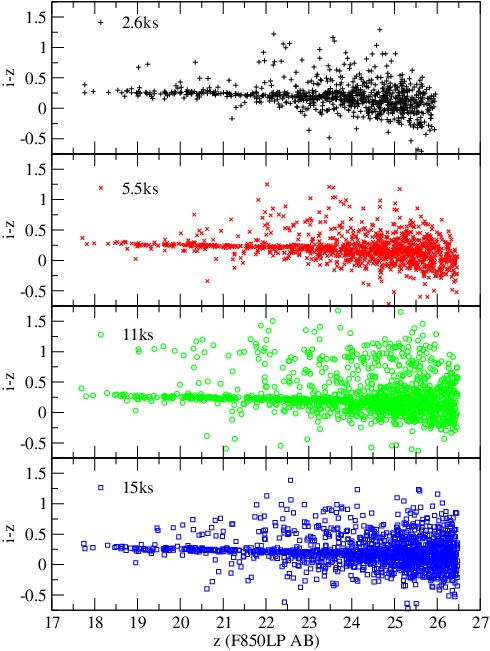} 
\caption{Colour-magnitude diagrams for galaxies in MS1358+62, each corresponding to one exposure
length. The colours are the same as in Figure~\ref{m1358sb}. From the top: 2.6ks, 5.5ks, 11ks and 14ks
(in the $z$ band only, the $i$ is 11ks).}
\vspace{1cm}
\label{cmdcomps}
\end{figure}

\begin{figure}
\includegraphics[width=0.475\textwidth]{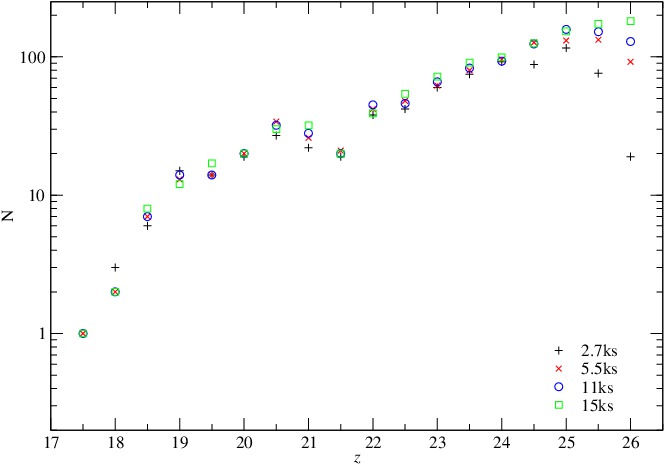} 
\caption{Red sequence luminosity distributions for galaxies in MS1358+62. Each colour and symbol corresponds
to a different exposure, as in Figures~\ref{m1358sb} -- ~\ref{cmdcomps}. These are also identified in the figure legend.
The figure demonstrates that objects are missed (and the luminosity function slope becomes flatter) because of
surface brightness selection effects, even if the nominal magnitude limit is exceeded.}
\label{lfsb}
\end{figure}

We plot the $i-z_{F850LP}$ vs. $z_{F850LP}$ colour-magnitude diagram for galaxies in all our exposures
(i.e., 2.6ks, 5.5ks and the combined 11ks and 15ks -- the latter in $z$ only) in Figure~\ref{cmdcomps} 
and fit single straight lines to the red sequence, as for all other clusters. 

We have then selected galaxies on the red sequence in $i-z_{F850LP}$ for a $z_{F850LP}$-selected sample.
Figure~\ref{lfsb} shows how the luminosity distribution for  galaxies on the red sequence is affected by the
exposure times (and therefore  surface brightness limit reached, as in Figure~\ref{m1358sb}). While the
brighter portion is unaffected (all galaxies  lie well above the selection limit) we see how the luminosity
distributions for  the fainter images start to deviate systematically from that found in the shorter (2.7ks)
exposure (which is  more typical of most of our data above, as well as previous work such as the EDisCS 
survey) at $z_{F850LP} > 24$ and that this also extends to points representing progressively longer  exposures (for
clarity we do not  show error bars here but it is apparent that at $z_{F850LP} > 23$ the deeper images contain more
galaxies). This  is not due to detection limit of the images, even for the 2.6ks exposures, which reach to 
$z_{F850LP} \sim 27$. Dwarf galaxies may lie at low surface brightness well within the normally computed
completeness limit, but go undetected because their central surface brightness is too low,  as early pointed 
out (in a different context)  by \cite{Disney1976} and in a similar (luminosity function-specific)
fashion by \cite{Phillipps1995} and \cite{Driver2003}.

We can follow the red sequence to $z_{F850LP} \sim 26$ at least {\it with no hint of a decrease} at faint luminosities. 
We  also note that more and more faint dwarfs emerge from the sky as the exposure length increases,
suggesting that  the census of dwarf galaxies in clusters may be incomplete. A fuller analysis of this and 
other clusters (with deep archival HST data in the Sloan  filters) will be presented in a separate paper.

\begin{figure}
\includegraphics[width=0.475\textwidth]{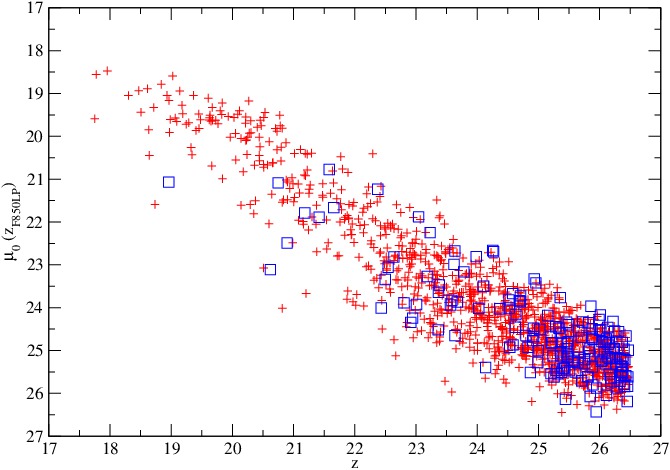} \\
\\
\\
\\
\includegraphics[width=0.475\textwidth]{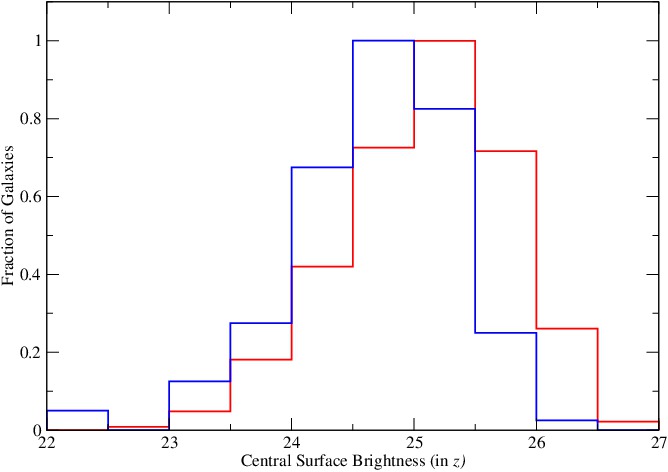} 
\caption{Top panel: Surface brightness distribution of blue cloud (blue squares) and red sequence
 (red crosses) in
MS1358+62. In the bottom panel we show a histogram of the central surface brightness distribution for red and blue 
galaxies for $z_{F850LP} > 24$; this demonstrates that red galaxies have a broader central surface brightness distribution,
extending to lower luminosities and therefore are more susceptible to selection effects.}
\label{brc}
\end{figure}

Figure~\ref{brc} compares the central (detection) surface brightness distributions for galaxies on the red sequence
and blue cloud. Blue cloud galaxies, because they are star-forming, have higher central surface brightnesses than
red cloud dwarfs and therefore the latter galaxies will be preferentially missed because of surface brightness 
selection effects. This is also shown as a histogram in the figure, comparing the central surface brightness distributions 
for red and blue galaxies with $z_{F850LP} > 24$.  

This suggests that the apparent evolution of the red sequence in the EDiscCS sample \citep{DeLucia2007,
Rudnick2009} might be explained by a combination of (a) the decline in the fraction of intermediate 
luminosity galaxies owing to the non-monotonic shape of the luminosity distribution; (b) the possible 
faster evolution of this feature (relative to the true giants around $M^*$ if it is produced by the quenching 
of fainter blue galaxies as they evolve on to the red sequence \citep{Peng2010} and (c) a combination of 
surface brightness selection effects and $(1+z)^4$ dimming, coupled with the less stable PSF and sky 
level of ground-based observations, as noted by \cite{Andreon2008} and \cite{Crawford2009}. In addition, 
the greater detectability of blue cloud dwarfs naturally explains their much weaker evolution. It is interesting 
to note how the apparent decline in the red sequence in Figure~\ref{lfsb} parallels the 'merger' induced
evolution in (e.g.,) Rudnick et al. (2012, their Figure 6) as well as the redshift evolution of the red sequence:
for instance, the shortest exposure (2.7ks) may be taken to be the equivalent of the more complete and
longest exposure (15ks) at $z=0.5$.

\section{Conclusions}

In this work we have sought to characterize the luminosity functions and colour-magnitude diagrams of
collisional galaxy clusters at redshifts $0.2 < z < 0.6$, and to compare these to the LFs of normal, non-colliding, clusters. We come to the following conclusions.

\begin{itemize}
\item Regardless of how far advanced is the merger or collision, there appears to be no difference between collisional and normal clusters in terms of their overall or red/blue separated luminosity functions. There is a similar lack of difference between the characteristics of their red sequences.

\item All overall cluster luminosity functions are consistent with passive evolution of the galaxy population and have a fixed faint end slope $\alpha \simeq -1$

\item We infer from this that the collisions have not significantly affected the properties of the cluster galaxy populations, even down to the 
low luminosity and low mass regime of dwarf galaxies.

\item Contrary to some previous work there is no evidence of weakening of the red sequence (at the faint end) in the  higher redshift members of our sample. We tentatively attribute this difference to increasing surface brightness driven incompleteness at faint
magnitudes in previous studies, as this preferentially removes faint red objects. However, we cannot rule out the possibility that the weakening is dependent on cluster mass, with our clusters typically having higher masses than those in previous studies. We note, though, that we see no difference in the red sequence even in the lower mass subclusters which make up our merging systems. 

\item Because the red sequences can be tracked down to faint luminosities and low masses ($<10^9 M_{\odot}$) this implies that red dwarf galaxies
were already formed prior to the epoch of observation and prior to any merger events (in the merging clusters) i.e. at $z > 0.7$ (7 Gyrs ago).

\item There is a distinct increase in the fraction of blue cluster galaxies at higher redshifts ($z \sim 0.6$), at all luminosities. This is more general than the 
classical Butcher-Oemler effect as this increase affects all galaxies in the blue cloud, not just the bluest objects. Bright ($M_I < -21$) blue galaxies are much more common in high redshift clusters than in low redshift systems. Whether this reflects evolution in the luminosity function 
of the infalling field galaxies or an increase in the quenching efficiency with decreasing redshift is unclear.
\end{itemize}

\section*{Acknowledgements}
This work has made use of the Hubble Legacy Archive (HLA) and the Mikulski Archive for the Space Telescope (MAST).
The PIs of the original projects which produced these data are thanked for providing excellent deep archival images with diverse uses beyond the original programmes. We would like to thank the anonymous referee for a constructive series of reports.

\bsp
\label{lastpage}
 

\newpage 
\begin{thebibliography}{99}
\bibitem[\protect\citeauthoryear{Andreon}{2006}]{Andreon2006a}
Andreon S. 2006, MNRAS, 369, 969
\bibitem[\protect\citeauthoryear{Andreon et al.}{2006}]{Andreon2006b}
Andreon S., Cuillandre J.-C., Puddu E., Mellier Y. 2006, MNRAS, 372, 60
\bibitem[\protect\citeauthoryear{Andreon}{2008}]{Andreon2008}
Andreon S. 2008, MNRAS, 386, 1045
\bibitem[\protect\citeauthoryear{Armstrong \& Kung}{1978}]{Armstrong1978}
Armstrong R. D., Kung M. T. 1978, Journal of the Royal Statistical Society
series C (Applied Statistics), 27, 363
\bibitem[\protect\citeauthoryear{Barkhouse et al.}{2007}]{Barkhouse2007}
Barkhouse W. A., Yee H. K. C., Lopez-Cruz O. 2007, ApJ, 671, 1471
\bibitem[\protect\citeauthoryear{Bell et al.}{2004}]{Bell2004}
Bell E. F. et al. 2004, ApJ, 608, 752
\bibitem[\protect\citeauthoryear{Bertin \& Arnouts}{1996}]{Bertin1996}
Bertin E., Arnouts S. 1996, A\&AS, 117, 393
\bibitem[\protect\citeauthoryear{Bildfell et al.}{2012}]{Bildfell2012}
Bildfell C. et al. 2012, MNRAS, 425, 204
\bibitem[\protect\citeauthoryear{Bourdin et al.}{2011}]{Bourdin2011}
Bourdin H. et al. 2011, A\&A, 527, A21
\bibitem[\protect\citeauthoryear{Boyce et al.}{2001}]{Boyce2001}
Boyce P. J., Phillipps S., Jones B., Driver S. P., Smith R. M., Couch W. J.
2001, MNRAS, 328, 277
\bibitem[\protect\citeauthoryear{Bradac et al.}{2008}]{Bradac2008} 
Bradac M., Allen S. W., Treu T., Ebeling H., Massey R., Morris R. G., von der Linden A., Applegate D.
2008, ApJ, 687, 959
\bibitem[\protect\citeauthoryear{Bruzual \& Charlot}{2003}]{Bruzual2003}
Bruzual G., Charlot S. 2003, MNRAS, 344, 1000
\bibitem[\protect\citeauthoryear{Butcher \& Oemler}{1978}]{Butcher1978}
Butcher H., Oemler A. 1978, ApJ, 226, 559
\bibitem[\protect\citeauthoryear{Butcher \& Oemler}{1984}]{Butcher1984}
Butcher H., Oemler A. 1984, ApJ, 285, 426
\bibitem[\protect\citeauthoryear{Clowe et al.}{2006}]{Clowe2006}
Clowe D., Bradac M., Gonzalez A. H., Markevitch M., Randall S. W., Jones C., Zaritsky D.
2006, ApJ, 648, L109
\bibitem[\protect\citeauthoryear{Clowe et al.}{2012}]{Clowe2012} 
Clowe D., Markevitch M., Bradac M., Gonzalez A. H., Chung S. M., Massey R., Zaritsky D.
2012, ApJ, 758, A128
\bibitem[\protect\citeauthoryear{Colless}{1989}]{Colless1989}
Colless M. M. 1989, MNRAS, 237, 799
\bibitem[\protect\citeauthoryear{Cowie et al.}{1996}]{Cowie1996}
Cowie L. L., Songaila A., Hu E. M., Cohen J. G. 1996, AJ, 112, 839
\bibitem[\protect\citeauthoryear{Crawford et al.}{2006}]{Crawford2006}
Crawford S. M., Bershady S., Glenn A. D., Hoessel J. G. 2006, ApJ, 636, L13
\bibitem[\protect\citeauthoryear{Crawford et al.}{2009}]{Crawford2009}
Crawford S. M., Bershady S., Hoessel J. G. 2009, ApJ, 690, 1158
\bibitem[\protect\citeauthoryear{Davidzon et al.}{2013}]{Davidzon2013}
Davidzon I. et al. 2013, astro-ph, 1303.3808
\bibitem[\protect\citeauthoryear{Dawson et al.}{2012}]{Dawson2012}
Dawson W. A. et al. 2012, ApJ, 747, L42
\bibitem[\protect\citeauthoryear{de Filippis et al.}{2011}]{DeFilippis2011}
de Filippis E., Paolillo M., Longo G., La Barbera F., de Carvalho R. R., Gal, R.
2011, MNRAS, 414, 2771
\bibitem[\protect\citeauthoryear{De Lucia et al.}{2007}]{DeLucia2007}
De Lucia G. et al. 2007, MNRAS, 374, 809
\bibitem[\protect\citeauthoryear{De Lucia \& Blaizot}{2007}]{DeLuciaBlaizot2007}
De Lucia G., Blaizot J. 2007, MNRAS, 375, 2
\bibitem[\protect\citeauthoryear{De Propris et al.}{1999}]{DePropris1999}
De Propris R., Stanford S. A., Eisenhardt P. R., Dickinson M., Elston R.
1999, AJ, 118, 719
\bibitem[\protect\citeauthoryear{De Propris et al.}{2003a}]{DePropris2003a}
De Propris R., Stanford S. A., Eisenhardt P. R., Dickinson M. 2003a, ApJ, 598, 20
\bibitem[\protect\citeauthoryear{De Propris et al.}{2003b}]{DePropris2003b}
De Propris R. et al. 2003b, MNRAS, 342, 725
\bibitem[\protect\citeauthoryear{De Propris et al.}{2004}]{DePropris2004}
De Propris R. et al. 2004, MNRAS, 351, 125
\bibitem[\protect\citeauthoryear{De Propris et al.}{2007}]{DePropris2007}
De Propris R., Stanford S. A., Eisenhardt P. R., Dickinson M., Rosati P. 
2007, AJ, 133, 2209
\bibitem[\protect\citeauthoryear{Disney}{1976}]{Disney1976}
Disney M. 1976, Nature, 263, 573
\bibitem[\protect\citeauthoryear{Driver et al.}{2003}]{Driver2003}
Driver S. P., Odewahn S. C., Echevarria L., Cohen S. H., Windhorst R. A., Phillipps S., Couch, W. J.
2003, AJ, 126, 2662
\bibitem[\protect\citeauthoryear{Drory \& Alvarez}{2008}]{Drory2008}
Drory N., Alvarez M. 2008, ApJ, 680, 41
\bibitem[\protect\citeauthoryear{Ebeling et al.}{2010}]{Ebeling2010}
Ebeling H., Edge A. C., Mantz A., Barrett E., Henry J. P., Ma C. J., van Speybroeck L.
2010, MNRAS, 407, 83
\bibitem[\protect\citeauthoryear{Eisenhardt et al.}{2007}]{Eisenhardt2007}
Eisenhardt P. R., De Propris R., Gonzalez A. H., Stanford S. A., Wang M., Dickinson M.
2007, ApJS, 169, 225
\bibitem[\protect\citeauthoryear{Faber et al.}{2007}]{Faber2007}
Faber S. M. et al. 2007, ApJ, 665, 265
\bibitem[\protect\citeauthoryear{Giavalisco et al.}{2004}]{Giavalisco2004}
Giavalisco M. et al. 2004, ApJ, 600, L93
\bibitem[\protect\citeauthoryear{Gilbank et al.}{2008}]{Gilbank2008}
Gilbank D. G., Yee H. K. C., Ellingson E., Gladders M. D., Loh Y.-S., Barrientos L. F., Barkhouse W. A.
2008, ApJ, 673, 724
\bibitem[\protect\citeauthoryear{Girardi et al.}{2008}]{Girardi2008} 
Girardi M., Barrena R., Boschin W., Ellingson E. 2008, A\&A, 491, 379
\bibitem[\protect\citeauthoryear{Grogin et al.}{2011}]{Grogin2011}
Grogin N. et al. 2011, ApJS, 197, A35
\bibitem[\protect\citeauthoryear{Grutzbauch et al.}{2012}]{Grutz2012}
Gr\"{u}tzbauch R., Bauer A.E., Jorgensen I., Varela J., 2012, MNRAS, 423, 3652
\bibitem[\protect\citeauthoryear{Haines et al.}{2009}]{Haines2009}
Haines C. P. et al. 2009, ApJ, 704, 126
\bibitem[\protect\citeauthoryear{Hammer et al.}{2010}]{Hammer2010}
Hammer D. et al. 2010, ApJS, 191, 143
\bibitem[\protect\citeauthoryear{Harsono \& De Propris}{2007}]{Harsono2007}
Harsono D., De Propris R. 2007, MNRAS, 386, 1036
\bibitem[\protect\citeauthoryear{Harsono \& De Propris}{2009}]{Harsono2009}
Harsono D., De Propris R. 2009, AJ, 137, 3091
\bibitem[\protect\citeauthoryear{Hinshaw et al.}{2012}]{Hinshaw2012}
Hinshaw G. et al. 2012, arXiv 1212.5226
\bibitem[\protect\citeauthoryear{Holden et al.}{2009}]{Holden2009}
Holden B. P. et al. 2009, ApJ, 670, 190
\bibitem[\protect\citeauthoryear{Holden et al.}{2010}]{Holden2010}
Holden B. P., van der Wel A., Kelson D. D., Franx M., Illingworth G. D. 2010, ApJ, 721, 714
\bibitem[\protect\citeauthoryear{Huang et al.}{1997}]{Huang1997}
Huang J.-S., Cowie L. L., Gardner J. P., Hu E. M., Songaila A., Wainscoat R. J.
1997, ApJ, 476, 12
\bibitem[\protect\citeauthoryear{Hsu et al.}{2012}]{Hsu2012}
Hsu L.-Y., Ebeling H., Richard J. 2012, MNRAS, in press
\bibitem[\protect\citeauthoryear{Koekemoer et al.}{2011}]{Koekemoer2011}
Koekemoer A. et al. 2011, ApJS 197, A36
\bibitem[\protect\citeauthoryear{Lemaux et al.}{2012}]{Lemaux2012}
Lemaux B. C. et al. 2012, ApJ, 745, A106
\bibitem[\protect\citeauthoryear{Lerchster et al.}{2011}]{Lerchster2011}
Lerchster M. et al. 2011, MNRAS, 411, 2667
\bibitem[\protect\citeauthoryear{Lewis et al.}{2002}]{Lewis2002}
Lewis I. J. et al. 2002, MNRAS, 334, 673
\bibitem[\protect\citeauthoryear{Li et al.}{2012}]{Li2012}
Li I. H. et al. 2012, ApJ, 747, A91
\bibitem[\protect\citeauthoryear{Limousin et al.}{2012}]{Limousin2012}
Limousin M. et al. 2012, A\&A, 544, A71
\bibitem[\protect\citeauthoryear{Loveday et al.}{2012}]{Loveday2012}
Loveday J. et al. 2012, MNRAS, 420, 1239
\bibitem[\protect\citeauthoryear{Lu et al.}{2009}]{Lu2009}
Lu T., Gilbank D. G., Balogh M. L., Bognat A. 2009, MNRAS, 399, 1858
\bibitem[\protect\citeauthoryear{Ma et al.}{2009}]{Ma2009}
Ma C.-J., Ebeling H., Barrett E. 2009, ApJ, 639, L56
\bibitem[\protect\citeauthoryear{Ma et al.}{2010}]{Ma2010}
Ma C.-J., Ebeling H., Marshall P., Shabbrack T. 2010, MNRAS, 406, 121 
\bibitem[\protect\citeauthoryear{Mancone et al.}{2010}]{Mancone2010}
Mancone C. L. et al. 2010, ApJ,  720, 284
\bibitem[\protect\citeauthoryear{Mancone et al.}{2012}]{Mancone2012}
Mancone C. L. et al. 2012, ApJ, 761, A141
\bibitem[\protect\citeauthoryear{Markevitch et al.}{2002}]{Markevitch2002}
Markevitch M. et al. 2002, ApJ, 567, L27
\bibitem[\protect\citeauthoryear{Markevitch et al.}{2004}]{Markevitch2004}
Markevitch M. et al. 2004, ApJ, 606, 819
\bibitem[\protect\citeauthoryear{Merten et al.}{2011}]{Merten2011}
Merten J. et al. 2011, MNRAS, 417, 333
\bibitem[\protect\citeauthoryear{Moustakas et al.}{2013}]{Moustakas2013}
Moustakas J. et al. 2013, ApJ, 767, A50
\bibitem[\protect\citeauthoryear{Muzzin et al.}{2008}]{Muzzin2008}
Muzzin A., Wilson G., Lacy M., Yee H. K. C., Stanford, S. A. 2008, ApJ, 686, 966
\bibitem[\protect\citeauthoryear{Owers et al.}{2011}]{Owers2011} 
Owers M. S., Randall S. M., Nulsen P. E. J., Couch W., David L. P., Kempner J. C.
2011, ApJ, 728, A27
\bibitem[\protect\citeauthoryear{Papovich et al.}{2010}]{Papovich2010}
Papovich C. et al. 2010, ApJ, 716, 1503
\bibitem[\protect\citeauthoryear{Peebles}{1975}]{Peebles1975}
Peebles P. J. E. 1975, ApJ, 196, 647
\bibitem[\protect\citeauthoryear{Peng et al.}{2010}]{Peng2010}
Peng, Y.-J. et al. 2010, ApJ, 721, 193
\bibitem[\protect\citeauthoryear{Perez-Gonzalez et al.}{2008}]{PerezGonzalez2008}
Perez-Gonzalez P. G. et al. 2008, ApJ, 675, 234
\bibitem[\protect\citeauthoryear{Phillipps \& Driver}{1995}]{Phillipps1995}
Phillipps S., Driver S. P. 1995, MNRAS, 274, 832
\bibitem[\protect\citeauthoryear{Pimbblet \& Jensen}{2012}]{Pimbblet2012}
Pimbblet K. A., Jensen P. C. 2012, arXiv, 1208.0667
\bibitem[\protect\citeauthoryear{Popesso et al.}{2006}]{Popesso2006}
Popesso P., Biviano A., B{\"o}hringer H., Romaniello M. 2006, A\&A, 445, 29
\bibitem[\protect\citeauthoryear{Pracy et al.}{2004}]{Pracy2004}
Pracy M. B., De Propris R., Driver S. P., Couch W. J., Nulsen P. E. J.
2004, MNRAS, 352, 1135
\bibitem[\protect\citeauthoryear{Quilis et al.}{2000}]{Quilis2000}
Quilis V., Moore B., Bower R. 2000, {\it Science}, 288, 1617
\bibitem[\protect\citeauthoryear{Ragozzine et al.}{2012}]{Ragozzine2012}  
Ragozzine B., Clowe D., Markevitch M., Gonzalez A. H., Bradac, M. 2012, ApJ, 744, A94
\bibitem[\protect\citeauthoryear{Raichoor \& Andreon}{2012}]{Raichoor2012}
Raichoor A., Andreon S. 2012, A\&A, 543, A19
\bibitem[\protect\citeauthoryear{Rudnick et al.}{2009}]{Rudnick2009}
Rudnick G. et al. 2009, ApJ, 700, 1559
\bibitem[\protect\citeauthoryear{Rudnick et al.}{2012}]{Rudnick2012}
Rudnick G., Tranh K.-V., Papovich C., Momcheva I., Willmer C. 2012, ApJ, 755, A14
\bibitem[\protect\citeauthoryear{Schechter}{1976}]{Schechter1976}
Schechter P. L. 1976, ApJ, 203, 297
\bibitem[\protect\citeauthoryear{Schlafly \& Finkbeiner}{2011}]{Schlafly2011}
Schlafly E., Finkbeiner D. P. 2011, ApJ, 737, A103
\bibitem[\protect\citeauthoryear{Skelton et al.}{2012}]{Skelton2012}
Skelton R. E., Bell E. F., Somerville R. S. 2012, ApJ, 753, A44
\bibitem[\protect\citeauthoryear{Smith et al.}{2009}]{Smith2009}
Smith R. J. et al. 2009, MNRAS, 392, 1265
\bibitem[\protect\citeauthoryear{Smith et al.}{2012}]{Smith2012}
Smith R. J., Lucey J. R., Price J., Hudson M. J.,  Phillipps S. 2012, MNRAS, 419, 3167
\bibitem[\protect\citeauthoryear{Soucail}{2012}]{Soucail2012} 
Soucail G. 2012, A\&A, 540, A61
\bibitem[\protect\citeauthoryear{Stott et al.}{2007}]{Stott2007}
Stott J. P., Smail I., Edge A. C., Ebeling H., Smith G. P., Kneib J-P., Pimbblet K. A. 2007, ApJ, 661, 95
\bibitem[\protect\citeauthoryear{Tajiri \& Kamaya}{2001}]{Tajiri2001}
Tajiri Y. Y., Kamaya H. 2001, ApJ, 562, L125
\bibitem[\protect\citeauthoryear{Tucker et al.}{1998}]{Tucker1998}
Tucker W. et al. 1996, ApJ, 496, L5
\bibitem[\protect\citeauthoryear{Weisz et al.}{2011}]{Weisz2011}
Weisz, D. R. et al. 2011, ApJ, 739, A5
\bibitem[\protect\citeauthoryear{Yamanoi et al.}{2012}]{Yamanoi2012}
Yamanoi H. et al. 2012, AJ, 144, A40
\bibitem[\protect\citeauthoryear{Zeimann et al.}{2012}]{Zeimann2012}
Zeimann G. R. et al. 2012, ApJ, 756, A115
\end{thebibliography}
\end{document}